\documentclass[longauth]{aa}
\usepackage{amsmath}
\usepackage{mathrsfs}
\usepackage{dblfloatfix}
\usepackage{color}
\usepackage{graphicx}
\usepackage{fixltx2e}
\usepackage{float}
\usepackage{txfonts}
\usepackage{placeins}
\usepackage{array, makecell} 
\usepackage{gensymb}
\usepackage{booktabs}
\usepackage{url}
\usepackage{subfloat}
\usepackage[version=3]{mhchem}
\usepackage{threeparttable}
\usepackage{multirow}
\usepackage{CJK}
\usepackage{longtable}
\usepackage{pdflscape}
\usepackage[export]{adjustbox}
\usepackage{subfig}
\usepackage{array}
\usepackage{bm}
\usepackage[singlelinecheck=false, labelfont=bf]{caption}

\usepackage{natbib,twoopt}
\usepackage[hyphenbreaks]{breakurl}
\usepackage[breaklinks]{hyperref}      
\bibpunct{(}{)}{;}{a}{}{,}             
\definecolor{cobalt}{rgb}{0.06, 0.2, 0.65}
\hypersetup{
  colorlinks,
  citecolor=cobalt,
  linkcolor=cobalt,
  urlcolor=cobalt,
}
\makeatletter
  \newcommandtwoopt{\citeads}[3][][]{\href{http://adsabs.harvard.edu/abs/#3}%
    {\def\hyper@linkstart##1##2{}%
     \let\hyper@linkend\@empty\citealp[#1][#2]{#3}}}
  \newcommandtwoopt{\citepads}[3][][]{\href{http://adsabs.harvard.edu/abs/#3}%
    {\def\hyper@linkstart##1##2{}%
     \let\hyper@linkend\@empty\citep[#1][#2]{#3}}}
  \newcommandtwoopt{\citetads}[3][][]{\href{http://adsabs.harvard.edu/abs/#3}%
    {\def\hyper@linkstart##1##2{}%
     \let\hyper@linkend\@empty\citet[#1][#2]{#3}}}
  \newcommandtwoopt{\citeyearads}[3][][]%
    {\href{http://adsabs.harvard.edu/abs/#3}
    {\def\hyper@linkstart##1##2{}%
     \let\hyper@linkend\@empty\citeyear[#1][#2]{#3}}}
\makeatother

\newcommand{\myemail}{\email{cyang@eso.org}}

\newcommand{\kms}{{\hbox {\,km\,s$^{-1}$}}}
\newcommand{\mum}{{\hbox {\,$\mu$m}}}
\newcommand{\kkmspc}{{\hbox {\,K\,km\,s$^{-1}$\,pc$^{2}$}}} 
\newcommand{\lsun}{{\hbox {$L_\odot$}}}
\newcommand{\msun}{{\hbox {$M_\odot$}}}

\newcommand{\hto}{{\hbox {H\textsubscript{2}O}}}

\newcommand{\htop}{{\hbox {H$_2$O$^+$}}}

\newcommand{\ihto}{\hbox {$I_{\mathrm{H_2O}}$}}

\newcommand{\lco}{\hbox {$L_{\mathrm{CO}}$}}

\newcommand{\lir}{\hbox {$L_{\mathrm{IR}}$}}

\label{key}
\newcommand{\td}{{\hbox {$T_{\mathrm{d}}$}}}

\def\co#1#2{{\hbox {${\mathrm{CO}}(#1\text{--}#2)$}}}

\def\lpcol#1#2{\hbox {$L^{\prime}_{\mathrm{CO}(#1\text{--}#2)}$}}

\def\lco#1#2{\hbox {$L_{\mathrm{CO}(#1\text{--}#2)}$}}

\def\htot#1#2#3#4#5#6{\hbox {\hto(\t#1#2#3#4#5#6)}}
\def\t#1#2#3#4#5#6{{\hbox {$#1_{#2#3}\text{--}#4_{#5#6}$}}}
\def\ihtot#1#2#3#4#5#6{\hbox {$I_{\mathrm{H_2O}(#1_{#2#3}\text{--}#4_{#5#6})}$}}
\def\lhtot#1#2#3#4#5#6{\hbox {$L_{\mathrm{H_2O}(#1_{#2#3}\text{--}#4_{#5#6})}$}}

\defcitealias{2013A&A...551A.115O}{O13}
\defcitealias{2013ApJ...771L..24Y}{Y13}
\defcitealias{2016A&A...595A..80Y}{Y16}

\newcommand{\hz}{\hbox {high-redshift}}

\newcommand{\hatlas}{\hbox {{\it H}-ATLAS}}
\newcommand{\FWHM}{\textnormal{FWHM}}

\begin{document}

\begin{CJK*}{UTF8}{gbsn}

\title{CO, H\textsubscript{2}O, H\textsubscript{2}O$^+$ line 
and dust emission in a \textbf{\textit{z}}\,=\,3.63 strongly lensed 
starburst merger at sub-kiloparsec scales}

\author 
{ 
 C. Yang\;(杨辰涛)\inst{1}         \and 
 R. Gavazzi\inst{2}               \and  
 A. Beelen\inst{3}                \and  
 P. Cox\inst{2}                   \and  
 A. Omont\inst{2}                 \and  
 M. D. Lehnert\inst{2}            \and \\ 
 Y. Gao\;(高煜)\inst{4}            \and 
 R. J. Ivison\inst{5,6}           \and 
 A. M. Swinbank\inst{7}           \and
 L. Barcos-Mu\~noz\inst{8,9}      \and 
 R. Neri\inst{10}                 \and \\ 
 A. Cooray\inst{11}               \and
 S. Dye\inst{12}                  \and
 S. Eales\inst{13}                \and
 H. Fu\;(付海)\inst{14}           \and
 E. Gonz{\'a}lez-Alfonso\inst{15} \and
 E. Ibar\inst{16}                 \and
 M. J. Micha\l{}owski\inst{17}    \and \\ 
 H. Nayyeri\inst{11}              \and 
 M. Negrello\inst{13}             \and
 J. Nightingale\inst{7}           \and 
 I. P\'{e}rez-Fournon\inst{18,19} \and 
 D. A. Riechers\inst{20}          \and 
 I. Smail\inst{7}                 \and 
 P. van der Werf\inst{21}         
}

\institute{
European Southern Observatory, Alonso de C{\'o}rdova 3107, Casilla 19001, 
Vitacura, Santiago, Chile. 
\myemail
\and
Institut d'Astrophysique de Paris, Sorbonne Universit\'e, CNRS,  UMR 7095, 
98 bis bd Arago, 75014 Paris, France.
\and
Institut d'Astrophysique Spatiale, CNRS UMR 8617, Universit\'{e} Paris-Sud, 
Universit\'{e} Paris-Saclay, 91405 Orsay, France
\and  
Purple Mountain Observatory/Key Lab of Radio Astronomy, Chinese Academy of 
Sciences, Nanjing 210034, PR China 
\and
European Southern Observatory, Karl Schwarzschild Stra{\ss}e 2, 85748, 
Garching, Germany 
\and
Institute for Astronomy, University of Edinburgh, Royal Observatory, Blackford 
Hill, Edinburgh, EH9 3HJ, UK
\and
Centre for Extragalactic Astronomy, Durham University, Department of Physics, 
South Road, Durham DH1 3LE, UK
\and
National Radio Astronomy Observatory, 520 Edgemont Road, Charlottesville, VA 22903, USA
\and
Joint ALMA Observatory, Alonso de C{\'o}rdova 3107, Vitacura, Santiago, Chile 
\and
Institut de Radioastronomie Millim{\'e}trique (IRAM), 300 rue de la Piscine, 
38406 Saint-Martin-d'H{\`e}res, France  
\and
Department of Physics and Astronomy, University of California, Irvine, 
CA 92697, USA
\and
School of Physics and Astronomy, University of Nottingham, University Park, Nottingham NG7 2RD, UK
\and
School of Physics and Astronomy, Cardiff University, The Parade, Cardiff CF24 3AA, UK
\and 
Department of Physics \& Astronomy, University of Iowa, Iowa City, IA 52245
\and
Universidad de Alcal{\'a}, Departamento de F\'{i}sica y Matem{\'a}ticas, 
Campus Universitario, 28871 Alcal{\'a} de Henares, Madrid, Spain
\and 
Instituto de F\'{i}sica y Astronom\'{i}a, Universidad de Valpara\'{i}so, 
Avda. Gran Breta\~na 1111, Valpara\'{i}so, Chile
\and 
Astronomical Observatory Institute, Faculty of Physics, Adam Mickiewicz University, 
ul.~S{\l}oneczna 36, 60-286 Pozna{\'n}, Poland
\and
Instituto de Astrof\'\i sica de Canarias, C/V\'\i a L\'actea, s/n, 
E-38205 San Crist\'obal de La Laguna, Tenerife, Spain
\and
Universidad de La Laguna, Dpto. Astrof\'\i sica, E-38206 San 
Crist\'obal de La Laguna, Tenerife, Spain
\and
Department of Astronomy, Cornell University, 
220 Space Sciences Building, Ithaca, NY 14853, USA
\and
Leiden Observatory, Leiden University, Post Office Box 9513, 
NL-2300 RA Leiden, The Netherlands
}

\date {Received .../ Accepted ...}


\abstract
{Using the Atacama Large Millimeter/submillimeter Array (ALMA), we report 
high angular-resolution observations of the redshift $z$\,=3.63 galaxy 
{\it H}-ATLAS\,J083051.0+013224 (G09v1.97), one of the most luminous 
strongly lensed galaxies discovered by the {\it Herschel}-Astrophysical 
Terahertz Large Area Survey ({\it H}-ATLAS). We present 0\farcs2--0\farcs4 
resolution images of the rest-frame 188 and 419\,\mum\ dust continuum and 
the \co65, \htot211202, and $J_\mathrm{up}$\,=\,2 \htop\ line emission. 
We also report the detection of H$_2^{18}$O(\t211202) in this source. 
The dust continuum and molecular gas emission are resolved into a nearly 
complete $\sim$\,1\farcs5 diameter Einstein ring plus a weaker image in 
the center, which is caused by a special dual deflector lensing 
configuration. The observed line profiles of the \co65, \htot211202, 
and $J_\mathrm{up}$\,=\,2 \htop\ lines are strikingly similar. In the 
source plane, we reconstruct the dust continuum images and the spectral 
cubes of the CO, \hto, and \htop\ line emission at sub-kiloparsec scales. 
The reconstructed dust emission in the source plane is dominated by a 
compact disk with an effective radius of $0.7\pm0.1$\,kpc plus an 
overlapping extended disk with a radius twice as large. While 
the average magnification for the dust continuum is $\mu$\,$\sim$\,10--11, 
the magnification of the line emission varies from 5 to 22 across 
different velocity components. The line emission of \co65, \htot211202,  
and \htop\ have similar spatial and kinematic distributions. The molecular 
gas and dust content reveal that G09v1.97 is a gas-rich major merger in its 
pre-coalescence phase, with a total molecular gas mass of $\sim$\,10$^{11}$\,\msun. 
Both of the merging companions are intrinsically ultra-luminous infrared 
galaxies (ULIRGs) with infrared luminosities \lir\ reaching $\gtrsim$\,$4\times 10^{12}$\,\lsun,
and the total \lir\ of G09v1.97 is $(1.4 \pm 0.7)\times 10^{13}$\,\lsun. 
The approaching southern galaxy (dominating from $V$\,=\,$-$400 to $-$150\,\kms\   
relative to the systemic velocity) shows no obvious kinematic structure with 
a semi-major half-light radius of $a_\mathrm{s}$\,=\,0.4\,kpc, while the receding galaxy 
(0 to 350\,\kms) resembles an $a_\mathrm{s}$\,=\,1.2\,kpc rotating disk. The 
two galaxies are separated by a projected distance of 1.3\,kpc, bridged by 
weak line emission ($-$150 to 0\,\kms) that is co-spatially located with 
the cold dust emission peak, suggesting a large amount of cold interstellar 
medium (ISM) in the interacting region. As one of the most luminous star-forming 
dusty \hz\ galaxies, G09v1.97 is an exceptional source for understanding 
the ISM in gas-rich starbursting major merging systems at high redshift.
}


\authorrunning{C. Yang et al.}

\titlerunning{ALMA imaging of a $z$\,=\,3.63 strongly lensed starburst galaxy}

\maketitle


\section{Introduction}
\label{section:intro}

Some of the most vigorous starbursts that have ever occurred are
found in the high-redshift, dust-obscured submillimeter (submm)
galaxies \citep[SMGs,][]{1997ApJ...490L...5S, 1998Natur.394..248B,
1998Natur.394..241H}, or dusty star-forming galaxies \citep[DSFGs, 
see e.g., reviews by][]{2002PhR...369..111B, 2014PhR...541...45C}. 
With total infrared (IR) luminosities integrated over the rest-frame 
8--1000\,\mum, \lir\,$\geq$\,10$^{12}$\,\lsun\,(to even $\geq$\,10$^{13}$\,\lsun\ 
for a few), SMGs reach the limit of ``maximum starbursts''
\citep{2014ApJ...784....9B} with star formation rates (SFRs) that 
can exceed 1000\,\msun\,yr$^{-1}$ \citep[e.g.,][]{2015ApJ...807..128S}. 
Their extreme star formation rates indicate that these starburst galaxies 
are in the critical phase of rapid stellar mass growth, presumably 
consuming their gas reservoir on timescales $\lesssim$\,100\,Myr. 
Such intense star formation seen in high-redshift SMGs 
is thought to be triggered by galaxy mergers or at least enhanced 
by interaction with neighboring galaxies \citep[e.g.,][]{2006ApJS..163....1H,
2008ApJ...680..246T, 2010ApJ...724..233E, 2000MNRAS.315..209I, 
2011ApJ...743..159H, 2012ApJ...757...23K, 2012MNRAS.425.1320I, 
2012ApJ...760...11H, 2013Natur.498..338F, 2017A&A...602A..54M, 
2018MNRAS.476.2278H, 2018arXiv181107800L, 2018ApJ...864L..11X}. 
This is consistent with $\Lambda$CDM (Lambda cold dark matter) simulations 
where merger rates are expected to increase with increasing redshift 
\citep[e.g.,][]{2009ApJ...701.2002G, 2010MNRAS.406.2267F, 
2015MNRAS.449...49R}. Nevertheless, some simulations predict that 
such intense star formation can also be produced through secular 
processes, which are driven by high gas fraction and instabilities 
in isolated clumpy disks at high redshift \citep[e.g.,][]{2009ApJ...703..785D, 
2010MNRAS.404.1355D}. Such a scenario is also supported by 
observations \citep[e.g.,][]{2013ApJ...768...74T, 2016ApJ...833..103H, 
2018A&A...615A..25J}.

Spectroscopic follow-up observations determined a median 
redshift of $z$\,$\sim$\,2.5 for the SMG population discovered 
at 870\,\mum\ \citep[e.g.,][]{2005ApJ...622..772C, 2017ApJ...840...78D}, 
showing that they participate in the peak of the cosmic star formation 
rate density \citep{2014ARA&A..52..415M} at $z$\,$\sim$\,2--3 
\citep[e.g.,][]{2005ApJ...622..772C, 2011ApJ...732..126M,
2013A&A...553A.132M, 2014MNRAS.438.1267S}. In fact, relatively 
bright submm sources with $S_\mathrm{850\,{\mu}m}$\,>\,$1$\,mJy 
contribute a significant fraction to the cosmic star formation 
rate at this epoch \citep[$\gtrsim$\,10\%, e.g.,][]{2005ApJ...632..169L,
2011ApJ...732..126M, 2013A&A...553A.132M, 2014MNRAS.438.1267S,
2017MNRAS.466..861D, 2017MNRAS.469..492M}. These \hz\ SMGs have
many properties consistent with being progenitors of the local
massive spheroidal galaxies \citep[e.g.,][]{1999ApJ...518..641L,
2014ApJ...788..125S, 2014ApJ...782...68T,2018ApJ...856..121G}. 
They play a critical role in our understanding of the history of 
cosmic star formation and the physical processes underlying the 
most extreme phases of galaxy formation and evolution, although
their nature remains hotly debated \citep[e.g.,][]{2008ApJ...688..972C,
2010MNRAS.404.1355D, 2015Natur.525..496N}. This ongoing debate is
sustained, at least in part, because the sensitivity and/or resolution
of current observations are insufficient to unravel a complete picture
of the complex physical conditions and spatial structure of 
their interstellar media (ISM) and of the processes that regulate
the vigorous star formation.

Gravitational lensing provides one means to study the properties 
of the gas and dust at high spatial resolution and signal-to-noise 
ratio (S/N) in high-redshift galaxies by boosting the apparent flux 
and magnifying the apparent solid angle \citep[e.g.,][]{2010Sci...330..800N, 
2010Natur.464..733S, 2011ApJ...733L..12R, 2011ApJ...732L..35C, 
2015A&A...577A..50D, 2018ApJ...863L..16M}. At submm wavelengths, 
strongly lensed SMG candidates can be efficiently selected by applying 
a simple flux cut to survey images at far-IR, submm, and millimeter (mm) 
wavelengths, e.g., $S_{500\,{\mathrm \mu m}}$\,$>$\,100\,mJy 
\citep{2007MNRAS.377.1557N, 2010Sci...330..800N, 2017MNRAS.465.3558N, 
2013ApJ...762...59W}. 
Although strongly lensed SMGs are rare, $\lesssim$\,0.3\,deg$^{-2}$, 
statistically significant samples have recently become available, thanks  
to extragalactic wide-area surveys such as the {\it Herschel}-Astrophysical 
Terahertz Large Area Survey \citep[\hatlas, where sources are selected 
at 500\,$\mu$m,][]{2010PASP..122..499E}, the {\it Herschel} Multi-tiered 
Extragalactic Survey \citep[{\it Her}MES, also selected at 
500\,$\mu$m,][]{2012MNRAS.424.1614O}, the South Pole Telescope 
\citep[SPT, 1.4\,mm selected,][]{2013Natur.495..344V}, and the 
{\it Planck} all-sky survey \citep[e.g.,][]{2015A&A...582A..30P, 
2015A&A...581A.105C}. These surveys have enabled the discovery 
and follow-up of hundreds of strongly lensed SMGs. 

One of the most direct ways to understand the nature of these 
star-bursting dusty SMGs is by studying the raw ingredients 
that fuel their star formation, namely the content of their ISM. 
Such follow-up studies of the lensed SMGs have become routine 
\citep[e.g.,][]{2011ApJ...740...63C, 2011ApJ...738..125G, 
2011A&A...530L...3O, 2011MNRAS.415.3473V, 2012ApJ...753..134F, 
2012ApJ...757..135L, 2013Natur.495..344V, 2013ApJ...779...67B, 
2013A&A...551A.115O, 2014A&A...568A..92M, 2015MNRAS.452.2258D, 
2015A&A...581A.105C, 2015ApJ...806L..17S, 2016MNRAS.457.4406A, 
2016ApJ...826..112S, 2016A&A...595A..80Y, 2017A&A...608A.144Y, 
2017ApJ...850..170O, 2017ApJ...837...12W, 2018A&A...615A.142A, 
2018MNRAS.474.3866H, 2018MNRAS.481...59Z, 2018A&A...620A.61C, 
2018Natur.553...51M, 2018Sci...361.1016S}. However, most of these 
studies are limited in spatial resolution and only investigate their 
globally averaged properties. Spatially resolved observations with 
angular resolution approaching the characteristic scales of star-forming 
regions are still rare \citep[e.g.,][]{2015ApJ...806L..17S, 
2017A&A...604A.117C, 2018MNRAS.476.4383D, 2018MNRAS.477.4380S, 
2018A&A...610A..53M}. In order to understand the detailed physical 
properties of \hz\ SMGs, especially their complex intrinsic structures, 
it is crucial to acquire high angular resolution images. From such data, 
fundamental information about gas and dust with different properties 
(e.g., density, temperature, optical depth, and mass) can be gained 
and related to the spatial and kinematical structures within 
individual sources.

However, high spatial-resolution observations of the ISM in SMGs 
remain a technical challenge, mostly due to their great distances.
Indeed, such observations require high-sensitivity, long-baseline 
interferometric observations at submm/mm wavelengths. At $z$\,=\,3.6, 
a 1\arcsec\ beam translates into a physical resolution of $\sim$\,7\,kpc, 
which is comparable to the typical total extent of cold molecular 
gas reservoir within SMGs \cite[e.g.,][]{2011MNRAS.412.1913I, 
2016ApJ...832...78I, 2011ApJ...742...11S, 2011ApJ...733L..11R, 
2013ApJ...765....6S, 2015MNRAS.448.1874T}, and $\gtrsim$\,5 times larger 
than the size of its star-forming dense warm gas regions 
\cite[e.g., $\sim$\,1\,kpc,][]{2008ApJ...680..246T, 2013Natur.496..329R, 
2015ApJ...811..124S, 2015ApJ...806L..17S, 2015ApJ...798L..18H, 
2015ApJ...799...81S, 2016ApJ...833..103H}. Reaching spatial resolutions 
below $\sim$\,0\farcs07 ($\sim$\,500\,pc physical resolution), which 
are needed to resolve the star-forming dense warm gas regions of \hz\ SMGs, 
remains challenging for current observing facilities. Nevertheless, 
the magnification provided by strong gravitational lensing can boost 
the angular spatial resolving power by typical factors of $\sim$\,2--5 
\citep[e.g.,][]{2013ApJ...779...25B, 2015ApJ...812...43B, 2016ApJ...826..112S}, 
enabling us, in such cases, to perform high angular resolution observations 
with reasonable on-source integration times.

One of the brightest strongly lensed \hz\ SMGs in the {\it H}-ATLAS fields, 
J083051.0$+$013224 (hereafter G09v1.97), at $z$\,=\,3.63 \citep[the redshift 
was firstly measured from blind CO detections, Riechers\,et\,al.\,in\,prep., 
see][]{2013ApJ...779...25B}, is an ideal source for high spatial resolution 
observations. With a magnification factor of $\mu$\,=\,$6.9\pm0.6$, estimated 
from 880\,$\mu$m dust continuum observations with the SMA \citep{2013ApJ...779...25B}, 
the effective sensitivity is boosted by one order of magnitude and the angular 
resolution by an average factor of $\sim$\,3. Using the 1\arcsec\ SMA 
observation \citep{2013ApJ...779...25B} of G09v1.97, a spatial resolution 
of down to $\sim$\,2\,kpc scales has been reached in the reconstructed source 
plane. G09v1.97 is intrinsically luminous in the far-infrared (far-IR) with an 
estimated intrinsic total IR luminosity of \lir\,$\sim$\,$2.3\times10^{13}\,\lsun$ 
and a star formation rate surface density of $\sim$700\,\msun\,yr$^{-1}$\,kpc$^{-2}$ 
(taking the SMA-measured half-light radius size of 0.9\,kpc).
The total molecular gas mass is estimated to be $(1.1\pm0.5)\times10^{11}$\,\msun\ (derived 
from \lpcol10/10$^{10}$\,=\,$10.0\pm4.4$\,\kkmspc\ 
using a CO-to-gas-mass conversion factor of 0.8\,$M_\odot$(\kkmspc)$^{-1}$), 
with a gas density $\log({n_\mathrm{H_2}}/\mathrm{cm^{-3}})$\,=\,$3.3\substack{+0.8\\-0.9}$
and a kinematic temperature $\log(T_\mathrm{kin}/\mathrm{K})$\,=\,$2.30\pm0.47$
\cite[]{2017A&A...608A.144Y}. 

Here we present ALMA observations of the dust continuum, CO, \hto, and \htop\ 
line emission in G09v1.97 at spatial resolutions of $\lesssim$\,0\farcs4. 
Based on a lens model, the spatial distribution and kinematical 
structures of the molecular gas can be derived at angular resolutions 
$\lesssim$\,0\farcs1 in the source plane. The physical properties traced by 
the molecular line and dust emission can thus be spatially resolved on 
sub-kiloparsec (sub-kpc) scales, helping us to gain insight of the \hz\ 
SMG population. The paper is organized as follows: the observations and 
results are reported in Section\,\ref{section:obs}; Section\,\ref{section:results} 
describes the properties of the continuum and emission line images and the 
characteristics of the molecular line spectra; the lens modeling is outlined 
in Section\,\ref{section:lens_model}; Section\,\ref{section:source_plane} 
discusses the properties of the molecular gas and dust continuum in the 
source plane, including the gas kinematics. Finally, concluding remarks 
are given in Section\,\ref{section:conclusions}.

Throughout this work, we adopt a spatially flat $\Lambda$CDM cosmology 
with $H_{0}=67.8\pm0.9\,{\rm km\,s^{-1}\,Mpc^{-1}}$, $\Omega_\mathrm{M}=0.308\pm0.012$ 
\citep{2016A&A...594A..13P}, with an angular-size scale of 7.4\,kpc/\arcsec  
at $z$\,=\,3.632. Using a Chabrier \citeyearpar{2003PASP..115..763C} 
initial mass function (IMF), the calibration of the far-IR star formation 
rate ($\mathit{SFR}$) from \citet{2012ARA&A..50..531K} gives a relation of  
$\mathit{SFR}$\,=\,$1.4 \times 10^{-10}$\,(\lir/\lsun)\,\msun\,yr$^{-1}$ 
\citep[see also][]{2011ApJ...737...67M}.

\section{Observations and Data Reduction}
\label{section:obs}

\setlength{\tabcolsep}{0.245em}
\begin{table*}[!htbp]
\tiny
\caption{ALMA observation log of G09v1.97.}
{\renewcommand{\arraystretch}{1.1}
\begin{tabular}{cccccccccccccccccc}
\toprule
Band                   &   Date       &         \multicolumn{3}{c}{Calibrator}    & $t_\mathrm{total}$ & $t_\mathrm{on}$ &                                           \multicolumn{5}{c}{Condition}                                                 &  SPW        &   Science Goal       & $\nu_\mathrm{sky}$ &  \multicolumn{2}{c}{Synthesis Beam}                  \\\cmidrule{3-5}\cmidrule{8-12}\cmidrule{16-17} 
                       &              &  Bandpass   &    Flux      &  Phase       &                    &                 & $N_\mathrm{ant}$ & $\sigma_\psi$      & $\overline{PWV}$ & $\widetilde{T}_\mathrm{sys}$ &          Baseline             &             &                      &                    &        Size                       &       PA         \\
                       &              &             &              &              &       (min)        &    (min)        &                  &  (deg)             & (mm)             &       (K)                    &              (m)              &             &                      & (GHz)              &        (\arcsec)                  &      ($^\circ$)       \\
\midrule                                                                                                                                                                                                                                                                                                                                 
                       &              &             &              &              &                    &                 &                  &                    &                  &                              & \multicolumn{1}{c|}{ }        &     2       &       \co65\         &   149.207          &   0.37$\times$0.35                &          64      \\
\multirow{2}{*}{4}     & 04-Aug-2016  &  J0750+1231 &  J0750+1231  & J0825+0309   &         59.9       &     36.8        &    39            &    59              &  1.25            &        60                    & \multicolumn{1}{c|}{15--1396} &     0       &    \htot211202\      &   161.946          &   0.38$\times$0.36                &\llap{$-$}74      \\
                       & 25-Aug-2016  &  J0750+1231 &  J0854+2006  & J0825+0309   &         68.7       &     36.8        &    38            &    28              &  0.65            &        55                    & \multicolumn{1}{c|}{15--1462} &     1       &       \htop\         &   160.124          &   0.40$\times$0.36                &     85           \\
                       &              &             &              &              &                    &                 &                  &                    &                  &                              & \multicolumn{1}{c|}{ }        &  0--3       & Continuum\rlap{$^a$} &   154.508          &   0.32$\times$0.28                &   \llap{$-$}82   \\
\midrule                                                                                                                                                                                                              
 7                     & 01-Sep-2015  &  J0739+0137 &  J0510+1800  & J0839+0104   &         31.4       &      3.1        &    34            &    46              &  0.29            &        94                    &                     15--1574  &     --      & Continuum            &   343.494          &   0.19$\times$0.12                &     60           \\
\bottomrule
\end{tabular}}
   \begin{tablenotes}[flushleft]
   \small
	 \item\textbf{Note:} 
	      The central observing coordinates (J2000) are RA 08:30:51.156 and 
          DEC $+$01:32:24.35 for the Band 4 observations and RA 08:30:51.040, 
          DEC $+$01:32:25.000 for the Band 7 observations.
	      $N_\mathrm{ant}$ is the number of antennas used during the observations.
	      $\sigma_\psi$ and $\smash{\overline{PWV}}$ is the phase RMS 	
	      and the mean precipitable water vapor during the observations, respectively.
	      $\smash{\widetilde{T}_\mathrm{sys}}$ shows the median system temperature. 
	      SPW, namely the spectral windows cover the frequencies ranges of 
          161.100--162.819\,GHz (SPW-0), 159.278--160.996\,GHz (SPW-1), 
          148.345--150.064\,GHz (SPW-2) and 147.146--148.865\,GHz (SPW-3).
          $\nu_\mathrm{sky}$ gives the sky frequencies of the line centers and the continuum.  
          $^{(a)}$ The 2\,mm continuum image was made by combining a 
          line-free spectral window SPW-3 and all the line free channels from SPW-0, SPW-1, and SPW-2.
          The resulted 2\,mm continuum data has a representative frequency of 154.508\,GHz.
   \end{tablenotes}
   \label{table:obs-log}
\end{table*}
\normalsize 

The $z$\,=\,3.632\footnote{Based on the observed central frequency of
the \co65\ line from this work. The redshift $z$\,=\,3.632 corresponds
to a luminosity distance of $D_\mathrm{L}$\,=\,32724\,Mpc.} strongly
lensed SMG, G09v1.97 was observed in the 2\,mm atmospheric window (Band
4) with the Atacama Large Millimeter/submillimeter Array (ALMA), in the
project ADS/JAO.ALMA\#2015.1.01320.S (PI: A.\,Omont). The observations
used four spectral windows (SPW) covering two observed
frequency ranges of 147.146--150.064\,GHz and 159.278--162.819\,GHz
(see Table\,\ref{table:obs-log} for details). Data were acquired
during two observing executions on 04-Aug-2016 and 25-Aug-2016,
using 39 and 38 12-m antennas, respectively. The observations were
performed with the ALMA C36-5 configuration, which provides baselines
from 15\,m up to about 1462\,m, resulting in angular resolutions of
0\farcs3--0\farcs4 (with a Briggs robust weighting parameter of 0.5). The on-source
integration time was 36.8\,min for each execution, amounting to a
total of 73.8\,min on-source time, with a total amount of time for 
additional overheads of 54.8\,min. The overheads include pointing, 
focusing, phase, flux density and bandpass calibrations. J0825+0309 
was used as the phase calibrator and J0750+1231 as the bandpass calibrator. 
The flux calibrators were J0750+1231 and J0854+2006. A typical ALMA 
calibration uncertainty of 5\% is adopted for the Band 4 data. 
J0839+0104 was also used as a check source\footnote{A check source, which is used to
check the quality of the phase, is usually a bright quasar with 
a high-quality VLBI position, close to both the science target 
and the phase calibrator.}. The total available 7.5 GHz bandwidth 
of Band 4 was divided into four SPWs (i.e., SPW-0, SPW-1, SPW-2 and SPW-3, 
see Table\,\ref{table:obs-log} for details), each 1875\,MHz wide, covering
the major targeted lines of G09v1.97, i.e., \co65\ at 149.207\,GHz,
para-\htot211202\ (\htot211202\ hereafter) at 161.946\,GHz and series
of \htop\ lines (\htop(\t202111)\,$_{(5/2-3/2)}$ at the rest-frequency of 
742.1\,GHz and \htop(\t211202)\,$_{(5/2-3/2)}$ at the rest-frequency 
of 742.3\,GHz) with a representative frequency of 160.124\,GHz. In each
spectral window, there are 128 frequency channels giving a resolution
of 15.6\,MHz ($\sim$\,30\,\kms). The dust continuum in Band 4 is measured
by combining all the line-free channels, resulting a representative
frequency at 154.508\,GHz corresponding to 1.94\,mm or $\sim$\,419\,\mum\
in the rest-frame. The weather conditions were good with low water
vapor and stable phase during the two observing sessions as summarized
in Table\,\ref{table:obs-log}. The root mean square (RMS) of the data
reaches 0.21\,mJy\,beam$^{-1}$ in a 50\,\kms\ channel width.

We have also included in this study ALMA Band\,7 continuum archive data 
of G09v1.97 centered at 343.494\,GHz (ADS/JAO.ALMA\#2013.1.00358.S, PI: S.\,Eales, 
for further discussion of this dataset, see \citealt{2018MNRAS.475.4939A}),
which allow us to have a better constrain of the lens model and perform 
a detailed analysis of the spatial distribution and properties of dust emission.
The observed frequency corresponds to a wavelength of 0.873\,mm, or $\sim$\,188\,\mum\ 
in the rest-frame. The observations were performed on 01-Sep-2015 in 
good weather conditions, with 3.1 minutes on source time and a maximum baseline 
of 1.6\,km, yielding a synthesis beam of 0\farcs19$\times$0\farcs12 
(PA\,=\,60$^\circ$). The bandpass, flux and phase calibrators were J0739+0137, 
J0510+1800, and J0839+0104, respectively. The absolute flux calibration 
uncertainty in Band 7 is 10\%. Table\,\ref{table:obs-log} 
summarizes the details of the observations.  

Both datasets were calibrated using the ALMA calibration pipelines, 
with only minor flagging required. The calibrated data were then imaged and 
\texttt{CLEAN}ed using \texttt{tclean} within \texttt{CASA}\footnote{Common 
Astronomy Software Applications \citep{2007ASPC..376..127M}, see \url{https://casa.nrao.edu} for more 
information.} version 5.1.1, with a Briggs robust weighting factor of 
$-$\,0.5 for the Band 4 dust continuum to generate \texttt{CLEAN}-component 
models. All the line free channels were combined by using the \texttt{MS-MFS} 
algorithm \citep{2011A&A...532A..71R} with multiple Taylor terms 
\texttt{nterms=2} during the \texttt{CLEAN} process. We then performed 
several iterations of phase-only self-calibration until the S/N stopped  
improving. The typical phase variations are within $\pm$\,50\,deg and 
change smoothly with time. Accordingly, the corresponding \texttt{gaincal} 
solutions were applied to the entire dataset. After subtracting the 
dust continuum for the line emission data cube, the datasets were 
then \texttt{CLEAN}ed with a Briggs robust weighting factor of $-$\,0.5 
for the dust continuum (using again \texttt{nterms=2}, combining all 
line-free channels), $-$\,0.2 for the CO emission, and \,0.5 for 
the \hto\ and \htop\ emission, considering the optimization between 
the synthesized beamsize and achieved S/N level of the \texttt{CLEAN}ed 
images. Similar procedures of self-calibration data reduction for the 
Band 7 dust continuum were also performed to maximize the S/N.

\section{Results}
\label{section:results}

\subsection{Continuum and Emission Line Images}
\label{section:results:g09-img}

\begin{figure*}[!htbp]
\centering
\includegraphics[scale=0.155]{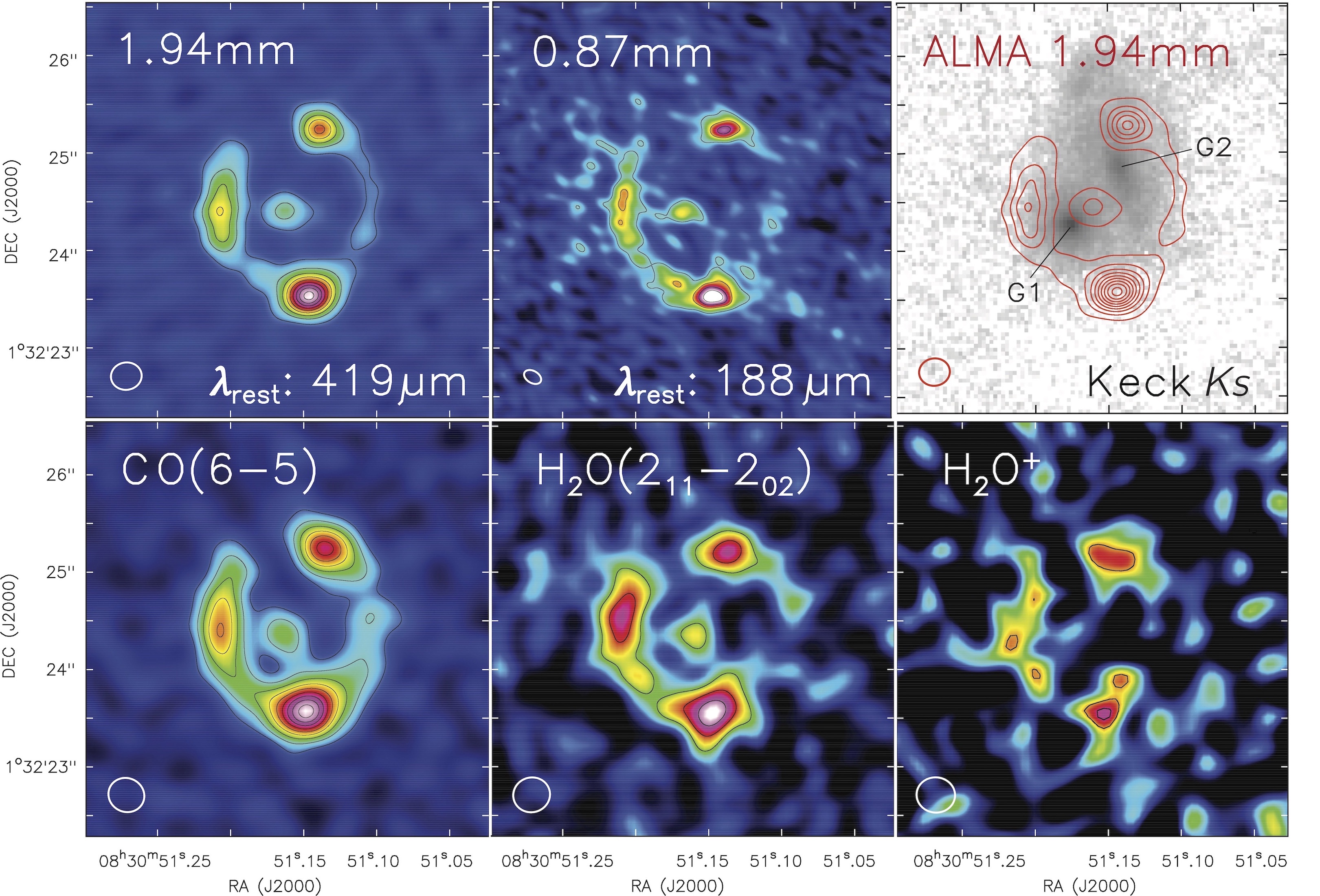}
\caption{ALMA images of the dust continuum and molecular gas emission
in G09v1.97.  {\it Upper row  - from left to right}: 1.94\,mm
(rest-frame $\sim$\,419\,\mum) dust continuum image with 0\farcs3 
resolution, with contours starting from $\pm$\,4\,$\sigma$ in steps of 
$\pm$\,6\,$\sigma$ ($\sigma$\,=\,0.03\,mJy\,beam$^{-1}$); 0.87\,mm 
(rest-frame $\sim$\,188\,\mum) dust continuum image with 
0\farcs19$\times$0\farcs12 resolution, with contours starting from 
$\pm$\,4\,$\sigma$ in steps of $\pm$\,6\,$\sigma$ 
($\sigma$\,=\,0.23\,mJy\,beam$^{-1}$); the 1.94\,mm dust continuum
contours overlaid on the {\it Keck}-II $\mathit{Ks}$-band image, which
shows the two deflecting foreground galaxies at $z$\,=\,0.626 (G1)
and $z$\,=\,1.002 (G2).  {\it Lower row - from left to right}:
Velocity-integrated molecular line emission images with 
0\farcs3--0\farcs4 resolution in: \co65\ with
contours from $\pm$\,4\,$\sigma$ in steps of $\pm$\,4\,$\sigma$;
\htot211202\ with contours from $\pm$\,4\,$\sigma$ in steps of
$\pm$\,3\,$\sigma$; and \htop\ with contours from $\pm$\,3\,$\sigma$
in steps $\pm$\,1\,$\sigma$. The $\sigma$ values for the CO, \hto,  
and \htop\ images are 0.05, 0.06, and 0.05\,Jy\,\kms\,beam$^{-1}$, 
respectively. The synthesized beam sizes are displayed in the lower 
left corners of each panel. The images show an almost complete 
Einstein ring, to first order, with similar spatial distributions 
for the dust continuum, CO and \hto\ line emission.  
}
\label{fig:g09-image}
\end{figure*}

\begin{figure*}[htbp]
\centering
\includegraphics[scale=0.56]{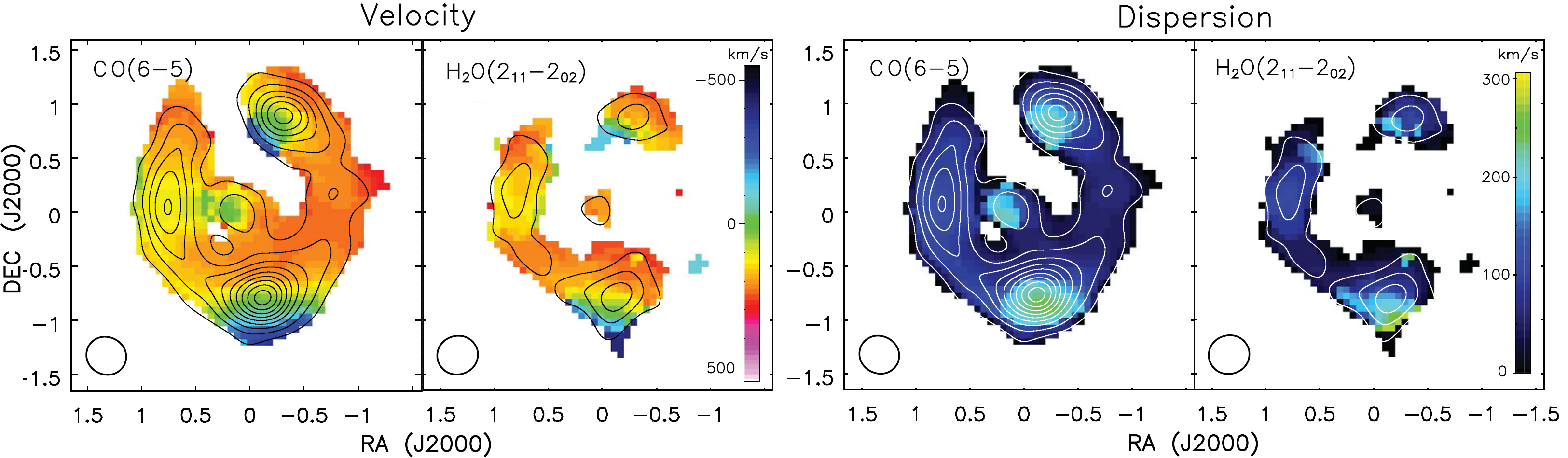}
\caption{
$\rm 1^{st}$ moment (velocity, left two panels) and $\rm 2^{nd}$ 
moment (velocity dispersion, right two panels) color maps of the \co65\ and 
\htot211202 line emission overlaid by the 0$\rm ^{th}$ moment 
(velocity-integrated line emission) contours. The images reveal the 
strongly lensed kinematic structure of G09v1.97 in the image plane 
in both \co65\ and \htot211202, with a significant velocity gradient 
seen in the northeastern and southern components. The CO and \hto\ 
lines trace similar kinematic structure as shown by the close 
correspondence between their first and second moment maps.}
\label{h2o:g09-mom-image}
\end{figure*}

The ALMA images of the dust continuum and the \co65, \htot211202, and 
\htop\ line emission of G09v1.97 are displayed in Fig.\,\ref{fig:g09-image}. 
In the upper row, the ALMA Band 4 dust continuum emission at 1.94\,mm 
($\sim$\,419\,\mum\ in the rest-frame), at a resolution of $\sim$\,0\farcs3, 
is shown next to the Band 7 dust continuum at 0.87\,mm (rest-frame 
$\sim$\,188\mum) at $\sim$\,0\farcs15. Both the dust continuum images 
show a very similar structure with a nearly complete $\sim$\,1\farcs5 
diameter Einstein ring, with three major image components. This is in 
agreement with the SMA\,880\,\mum\ dust continuum image \citep{2013ApJ...779...25B}. 
In addition, there is a weaker and smaller image component at the center, 
which was undetected with the SMA. Among the three image components, 
the one to the south is by far the brightest. Finally, there is extended 
emission which connects the southern and northern components along the 
eastern side of the ring. 

We also show the 1.94\,mm dust continuum superimposed on the {\it Keck}-II/NIRC2  
$\mathit{Ks}$-band image in Fig.\,\ref{fig:g09-image}. The figure shows 
the two foreground deflecting galaxies, the southern galaxy at $z$\,=\,0.626 (G1) 
and the northern one at $z$\,=\,1.002 (G2), with the redshifts obtained 
from the William Herschel Telescope (WHT) ACAM spectroscopy through the
detection of the Mg absorption and the [O{\footnotesize{\,II}}]$\lambda\lambda$3726,3729 
doublet lines, respectively. A {\it Hubble Space Telescope} ({\it HST}) 
image taken in the F110W band, which was obtained as part of the ID 12488 
Snapshot program (PI: Negrello), also confirms such a compound lens 
configuration.\footnote{Before comparing the ALMA images with the one 
from {\it HST}, we have corrected the registration of the archival 
optical/near-IR dataset with 9 Gaia stars in the field to ensure a good 
relative astrometry to within $0\farcs1$.} The unusual line-of-sight 
configuration, with two deflecting galaxies, complicates the model of 
the gravitational potential (Section\,\ref{section:lens_model}). This 
produces a central image that is not too de-magnified which is rarely 
seen in lensing configurations involving a single deflector with a 
cuspy mass distribution. 

G09v1.97 is neither detected in the {\it Keck}-II $\mathit{Ks}$-band 
(rest-frame 463\,nm) nor in the {\it HST} F110W-band (rest-frame 221\,nm) 
image, by checking the images after subtracting the foreground 
galaxies with \texttt{GALFIT} \citep{2002AJ....124..266P}. Comparing 
the rest-frame 419\,\mum\ dust continuum contours with the {\it Keck}-II 
image, it is evident that the contribution to the dust continuum from 
the two foreground deflectors is negligible. Finally, the similarities of 
the redshift and the profiles of the molecular line emission of the 
central component with the components along the Einstein ring rules 
out that this central emission is related to the deflecting galaxies.

Fig.\,\ref{fig:g09-image} also shows the continuum-subtracted images
with $\sim$\,0\farcs3--0\farcs4 resolution of the velocity-integrated
molecular line emission of \co65, \htot211202, and \htop\ (integrated
over the rest-frame 742.1\,GHz \htop(\t202111)\,$_{(5/2-3/2)}$ line 
in SPW-1 and the 746.5\,GHz \htop(\t211202)\,$_{(5/2-5/2)}$ line 
in SPW-0). Taking into account the range in S/N, all three molecular
gas lines display a nearly complete Einstein ring morphology, akin to
the one seen in the dust continuum emission, with a dominant sub-image 
in the south. The weak central image is detected in \co65\ and \htot211202\ 
but not in \htop. Nevertheless, the upper-limit from the \htop\ image is 
consistent with the flux ratios between the three sub-images and the 
central component found in the dust continuum, the \co65\ and \htot211202\ 
line images. By comparing the detailed differences between the line 
emission and the dust continuum, we find that the \co65\ line emission 
has a more extended morphology and resembles a more complete ring-like 
structure compared to the dust emission. Comparing the ratios between 
the brightest southern component and the two weaker ones in the northwest and northeast, 
the dust continuum of the southern component is about twice as bright, 
while for the \co65\ line emission, the ratio ranges from 1.3 to 1.7 
($\la$20\% uncertainties), and about 1.5 (with a larger uncertainty) 
for the \hto\ emission. This indicates that the rest-frame 
$\sim$\,200--400\,$\mu$m dust emission is slightly more concentrated
than the CO and \hto\ line emission. Finally, the \htop\ emission is
detected in the brightest southern component at $\sim$\,4-$\sigma$,
while it is only marginally detected in the two other components with
S/N ratios of about 3 (Fig.\,\ref{fig:g09-image}).

\begin{figure}[htbp]
\centering
\includegraphics[width=0.476\textwidth]{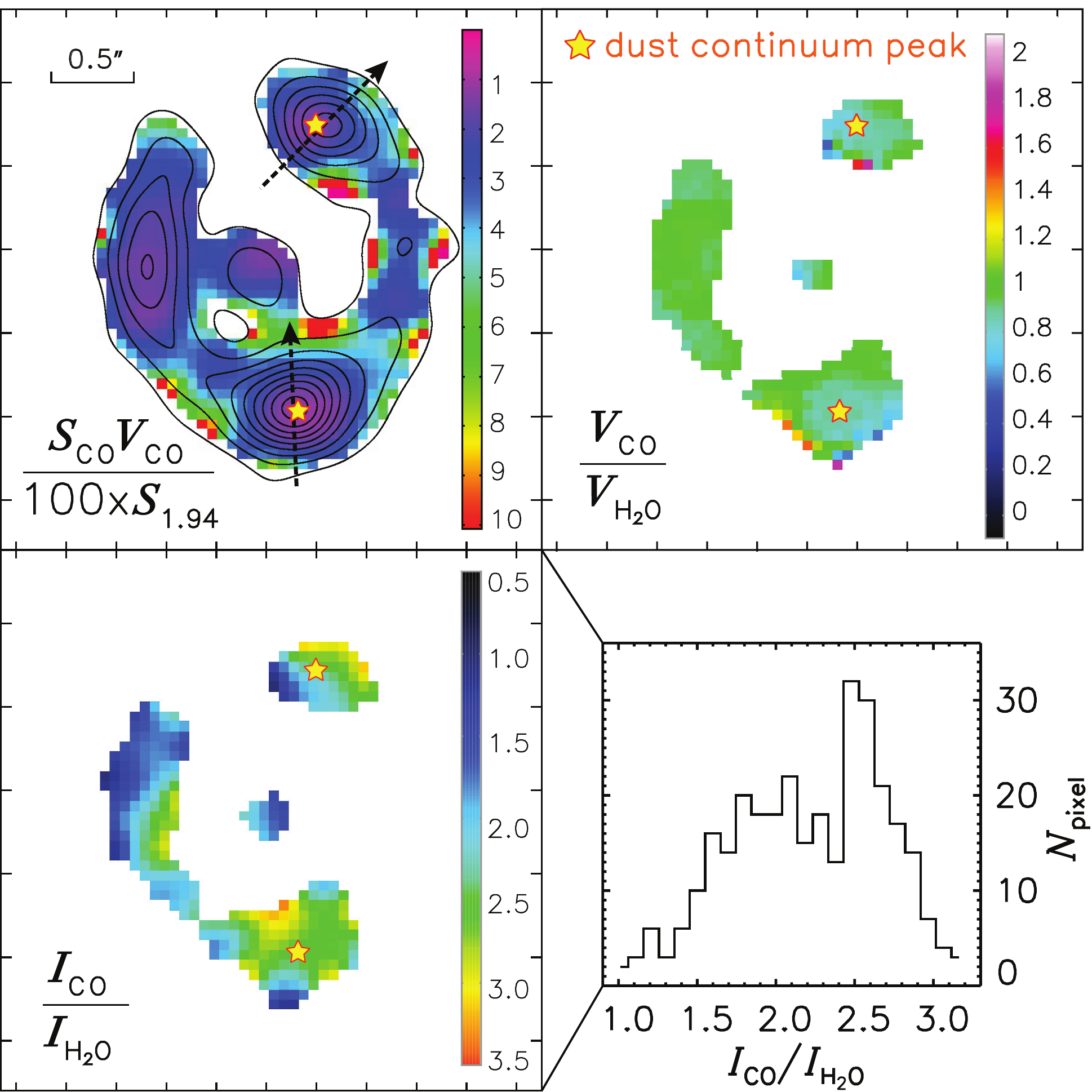}
\caption{
{\it Top left}: 
Ratio of the $\rm 0^{th}$ moment of \co65\ to  
the rest-frame 419\,$\mu$m (observed wavelength of 1.94\,mm) 
continuum. The contours are for the \co65\ $\rm 0^{th}$ moment and 
are the same as in Fig.\,\ref{h2o:g09-mom-image}.
{\it Top right}: 
$\rm 1^{st}$ moment ratio map of \co65\ to \htot211202.
{\it Bottom left}:
$\rm 0^{th}$ moment ratio map of \co65\ to \htot211202.
{\it Bottom right}:
Histogram of the \co65-to-\htot211202\ $\rm 0^{th}$ moment ratio.
The positions of the dust peak at north and south are indicated 
with yellow stars. The figure shows that the \co65\ emission is 
more extended than the dust continuum emission. Despite the small 
variation of the $\rm 0^{th}$ moment ratio, \co65 and \htot211202\ 
show very similar velocity structure.
}
\label{fig:mom1_ratio}
\end{figure}

Our three-dimensional ALMA data cubes allow us to further
study the kinematics of the \co65\ and \htot211202\ emitting
gas. Fig.\,\ref{h2o:g09-mom-image} shows the moment maps of the
\texttt{CLEAN}ed \co65\ and \htot211202\ data cubes. Both the moment
maps of CO and \hto\ show very similar distributions in velocity and
dispersion, although comparing the moments maps of \hto\ with the other
lines is hindered by the lower S/N. The moment maps reveal noticeable
velocity gradients in its major southern and northern components as
shown in the 1$\rm^{st}$ moment maps and are possibly arising from the
same lensed structure from the source plane. The velocity dispersion 
distributions compared to the 1$\rm^{st}$ moment maps show similar 
structure among the three image components (and only for two components 
in \htot211202\ due to the low S/N in the central component), with the 
peaks in velocity dispersion being slightly spatially offset 
from the continuum flux peaks. 

To better compare the dust continuum and the \co65\ 
and \htot211202\ line emission, in Fig\,\ref{fig:mom1_ratio} we show 
the ratio of the dust continuum to the 0$\rm^{th}$ moment of 
\co65, as well as the \co65-to-\htot211202\ ratio of the 0$\rm^{th}$ 
and 1$\rm^{st}$ moments. The map of $S_\mathrm{CO}V_\mathrm{CO}/S_\mathrm{1.94}$ 
shows clear evidence that the emission of \co65\ is more extended  
than the dust continuum at rest-frame 419\,$\mu$m. And there 
is indeed an offset between the dust and CO/\hto\ emission peaks.
The difference in sizes is consistent with several previous 
observations that the size of dust continuum is usually found 
smaller than the gas tracers such as the low- and mid-$J$ CO lines 
\citep[e.g.,][]{2011ApJ...733L..12R, 2011MNRAS.412.1913I, 
2015ApJ...811..124S, 2017ApJ...846..108C, 2018ApJ...863...56C, 
2018arXiv181012307H}. 
This could be caused by a radial dust temperature gradient 
\citep[see, e.g.,][]{2017A&A...602A..54M}. 
The 0$\rm^{th}$ moment ratio of \co65\ to \htot211202\ 
shows small variations, having flux ratios between \co65\ and 
\htot211202\ $\sim$\,1.7--2.8 with an average value of 2.4. 
The velocity structure of the two gas tracers are almost identical, 
with a ratio $V_\mathrm{CO}/V_\mathrm{H_2O}$\,=\,$1.0\pm0.2$.

\subsection{Integrated Spectra}
\label{section:g09-spec}

\begin{figure*}[!hbtp]
\centering
\includegraphics[width=0.8\textwidth]{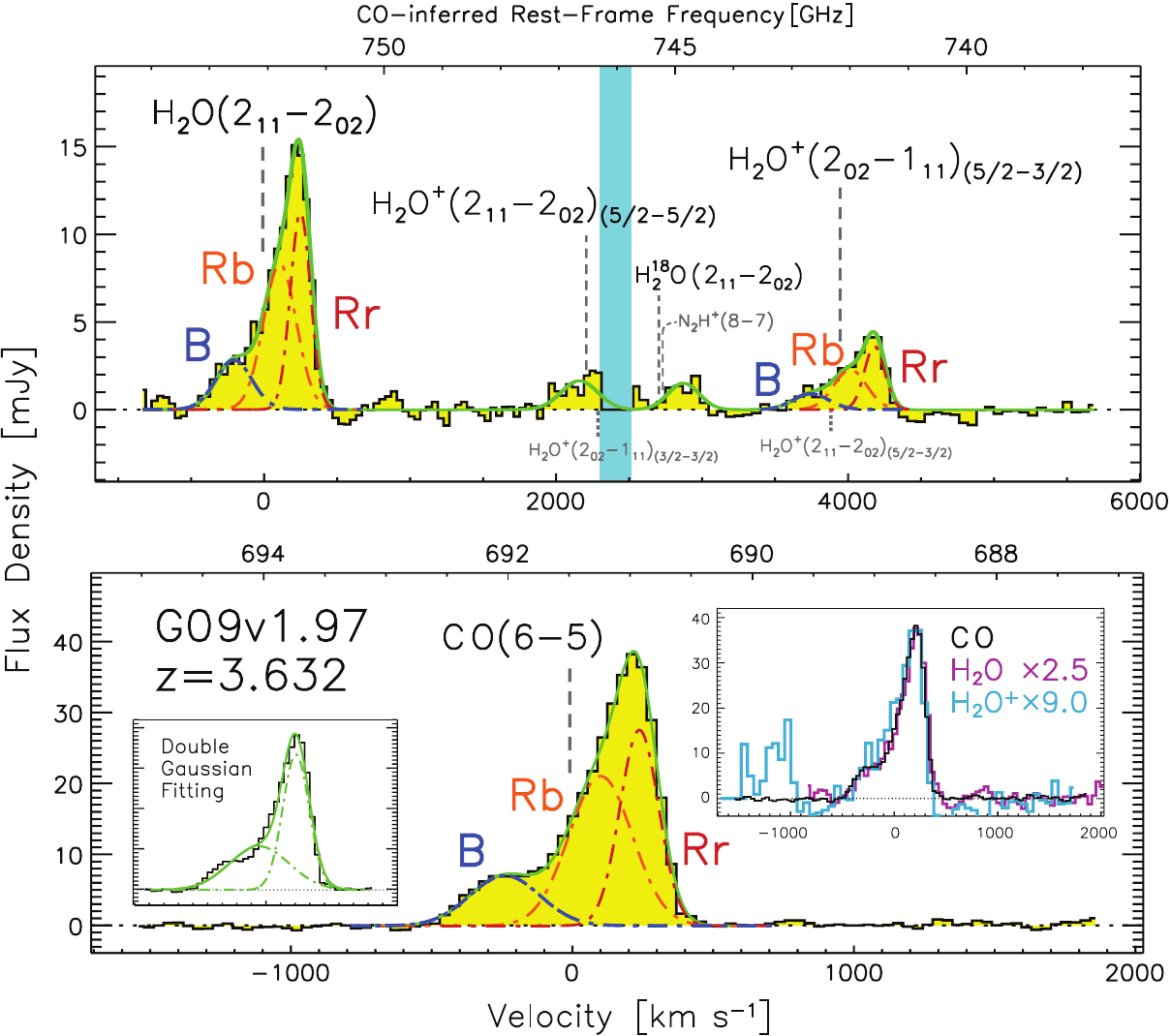}
\vspace{-0.2cm}
\caption{
Spatially integrated, continuum-subtracted 2\,mm spectra of G09v1.97 
with a spectral resolution of 35\,\kms. The vertical dotted 
lines represent the expected central positions of the lines. 
As argued in the text, that the weak emission at $\sim$\,745\,GHz 
feature is likely dominated by the H$_2^{18}$O(\t211202) line.  
The dot-dashed lines represent the corresponding Gaussian 
decompositions: blue represents the approaching gas component (marked 
as ``B'') while orange and red represent the blue-shifted Gaussian 
component (marked as ``Rb'') and red-shifted Gaussian component 
(marked as ``Rr'') of the receding gas, respectively. The solid 
green line shows the sum. The average noise level of the spectra 
is $\sim$\,0.2\,mJy with a spectral resolution of 50\,km\,s$^{-1}$.         
{\it Top}: Rest-frame 738--755\,GHz spectrum combining SPW-0 and 
SPW-1. The light-blue shaded area indicates the 746--746.5\,GHz gap 
between SPW-0 and SPW-1. The H$_2^{18}$O(\t211202) line and the 
\htop(\t202111)$_{(5/2\text{--}5/2)}$ line are fitted with a single 
Gaussian profile due to limited S/N and the lack of data in the 
inter-band gap.
{\it Bottom}: SPW-2 spectrum of \co65. The dot-dashed and solid 
lines are the same as in the top panel. The left inset shows 
the fit of the spectrum using two Gaussian profiles (green lines). 
It is clear that a double-Gaussian profile does not fit well the 
total line profile of the source. It is because of this, we 
introduced a third Gaussian component in the fit. The right inset 
shows a comparison among the line profiles of \co65, \htot211202, and 
\htop(\t202111)$_{(5/2\text{--}3/2)}$. They are found to be 
similar, indicating a similar spatial distribution of the 
line-emitting regions.
}
 \label{fig:img:g09-spec}
 \end{figure*}

The continuum-subtracted spectrum integrated over the entire source is 
shown in Fig.\,\ref{fig:img:g09-spec}. The upper panel shows the 
combined spectra of the 2\,mm windows SPW-0 and SPW-1 of the \hto\ and 
\htop\ lines, while the lower shows the \co65\ line covered by SPW-2. In 
addition to the strong \co65\ and \hto\ emission lines, there is a series 
of \htop\ emission lines including the dominant \htop\ feature, i.e.,   
\htop(\t211202)\,$_{(5/2-5/2)}$ and \htop(\t202111)\,$_{(5/2-3/2)}$ 
(based on the analysis of the expected relative strengths of the \htop\ 
submm lines in Arp\,220 by \citealt{2013A&A...550A..25G}). We also detect 
an emission line at 745\,GHz, which will be discussed at the end of 
this section.

After extracting the spectra integrated over the entire spatial region 
of the source, the emission lines were fitted with multiple Gaussian 
profiles using the Levenberg-Marquardt least-square minimization code 
\texttt{MPFIT} \citep{2009ASPC..411..251M}. Initially, two 
Gaussian components were fitted. However, as indicated in 
Fig.\,\ref{fig:img:g09-spec}, fitting the profile with two Gaussians 
results in significant residuals in the blue part of the line profiles. 
Therefore, we fitted the lines with three Gaussian components. 
The overall line profile of all 
the emission lines are well fitted by the three Gaussian components which we 
mark with ``B'', ``Rb'', and ``Rr'' in Fig.\,\ref{fig:img:g09-spec}. Since 
the line profiles of the \co65, \hto, and \htop\ lines agree very 
well with each other (Fig.\,\ref{fig:img:g09-spec}), we fix the linewidths of 
the B, Rb, and Rr components to be that found for the components of the 
\co65 line and use these line widths when fitting the profiles of the \hto\ 
and \htop\ lines. The results obtained, i.e., the velocity integrated 
line fluxes, linewidths (\FWHM) and line centroid positions are given in 
(Table\,\ref{table:obs-results}). The total integrated line 
fluxes for \co65\ and \htot211202\ are comparable to those obtained 
with the IRAM 30-meter telescope and NOEMA \citep{2016A&A...595A..80Y, 
2017A&A...608A.144Y} indicating that there is no missing 
flux in the ALMA data of G09v1.97. This implies that there is no significant diffuse 
emission (compared with the synthesis beam size) from either the dust 
continuum or any of the lines studied here.

\setlength{\tabcolsep}{0.275em}
\begin{table*}[!hbtp]
\small
\caption{Molecular line properties derived from the integrated spectra of G09v1.97.}
\begin{tabular}{lllrrrclcrc}
\toprule
 Line                                                            &  $\nu_\mathrm{center}$  &  Comp\rlap{.}      &$S_\mathrm{pk}$\;\;\;\;     &$S_\mathrm{Rr}/S_\mathrm{B}$\;\;& $I_\mathrm{line}$\,\;\;\;\; & $I_\mathrm{Rb+Rr}/I_\mathrm{B}$ & \;\;$\Delta{V}_\mathrm{line}$ & $\Delta{V}_\mathrm{Rb+Rr}$/$\Delta{V}_\mathrm{B}$  & \;$\mu L_\mathrm{Line}/10^{8}$ & $\mu L'_\mathrm{Line}/10^{10}$ \\
                                                                 &    (GHz)                &                    &     (mJy)\;\;\;            &                                & (Jy\,km\,s$^{-1}$)          &                                 & (km\,s$^{-1}$)                &                                                    &       (\lsun)\;\;\;\;          &       (\kkmspc)                \\
\midrule                                                                                                                                                                                                                                                                                                                                                          
\multirow{3}{*}{\co65}                                           &\multirow{3}{*}{149.287} & \;\;B             &\;\;\;   $7.1\pm0.8$        &\multirow{3}{*}{$3.9\pm1.1$}    &     $2.2\pm0.2$             &\multirow{3}{*}{$4.9\pm0.9$}     &  $293\pm21$                   &\multirow{3}{*}{$1.5\pm0.2$}                        &     $3.7\pm0.3$\;              &       $3.5\pm0.3$              \\
                                                                 &                         & \;\;Rb            &\;\;\;  $21.3\pm5.3$        &                                &     $6.0\pm1.3$             &                                 &  $265\pm33$                   &                                                    &    $10.0\pm2.2$\;              &       $9.4\pm2.0$              \\
                                                                 &                         & \;\;Rr            &\;\;\;  $27.7\pm7.1$        &                                &     $4.8\pm1.2$             &                                 &  $163\pm10$                   &                                                    &     $8.0\pm2.0$\;              &       $7.5\pm1.9$              \\[.2cm]
\multirow{3}{*}{\htot211202}                                     &\multirow{3}{*}{162.362} & \;\;B             &\;\;\;   $2.9\pm0.3$        &\multirow{3}{*}{$3.8\pm0.4$}    &     $0.9\pm0.1$             &\multirow{3}{*}{$4.7\pm0.5$}     &  $293$                        &      --                                            &     $1.6\pm0.2$\;              &       $1.2\pm0.1$              \\
                                                                 &                         & \;\;Rb            &\;\;\;   $8.2\pm0.4$        &                                &     $2.3\pm0.1$             &                                 &  $265$                        &                                                    &     $4.2\pm0.2$\;              &       $3.1\pm0.1$              \\
                                                                 &                         & \;\;Rr            &\;\;\;  $11.0\pm0.6$        &                                &     $1.9\pm0.1$             &                                 &  $163$                        &                                                    &     $3.4\pm0.2$\;              &       $2.5\pm0.1$              \\[.2cm]
\multirow{3}{*}{\llap{$^a$}\hto$^{+}$(\t202111)\,$_{(5/2-3/2)}$} &\multirow{3}{*}{161.2}   & \;\;B             &\;\;\;   $1.0\pm0.3$        &\multirow{3}{*}{$3.5\pm1.2$}    &     $0.3\pm0.1$             &\multirow{3}{*}{$4.0\pm1.4$}     &  $293$                        &      --                                            &     $0.5\pm0.2$\;              &       $0.4\pm0.1$              \\
                                                                 &                         & \;\;Rb            &\;\;\;   $2.1\pm0.4$        &                                &     $0.6\pm0.1$             &                                 &  $265$                        &                                                    &     $1.1\pm0.2$\;              &       $0.8\pm0.1$              \\
                                                                 &                         & \;\;Rr            &\;\;\;   $3.5\pm0.6$        &                                &     $0.6\pm0.1$             &                                 &  $163$                        &                                                    &     $1.1\pm0.2$\;              &       $0.8\pm0.1$              \\[.2cm]
\llap{$^b$}\hto$^{+}$(\t211202)\,$_{(5/2-5/2)}$                  &     160.2               & \;\;\llap{$^b$}B  &\;\;\;\llap{$^b$}$1.8\pm0.4$&      --                        & \llap{$^b$}$0.6\pm0.1$      &      --                         & \llap{$^b$}$294\pm43$         &      --                                            &     $1.1\pm0.2$\;              &       $0.8\pm0.1$              \\[.2cm]
\llap{$^c$}H$_2^{18}$O(\t211202)                                 &\llap{$^c$}160.907       & \;\;\llap{$^d$}Rr+Rb&\;\;\;   $1.7\pm0.5$        &      --                        &     $0.5\pm0.1$             &      --                         &   $250\pm58$                  &      --                                            &     $0.9\pm0.2$\;              &       $0.7\pm0.1$              \\
\midrule                                                                                                                                               
Dust Continuum  &     154.508          &  & 1.940\rlap{\,mm continuum}        &\multicolumn{7}{c}{\,$8.8\pm0.5$\,mJy}  \\
                &     343.494          &  & 0.873\rlap{\,mm continuum}        &\multicolumn{7}{c}{$106.6\pm10.7$\,mJy}  \\
\bottomrule
\end{tabular}
\tablefoot{ $\nu_\mathrm{center}$ is the sky frequency of the line center. $S_\mathrm{pk}$ is the 
	                      peak flux of the line component (Fig.\,\ref{fig:img:g09-spec}). 
                          Note that the linewidth has been fixed when fitting the \hto(\t211202)
                          and \htop(\t202111)$_{({5/2}\text{--}{3/2})}$ lines. The linewidths used in these fits were assumed to be the same as those determined from fitting the CO line profile.                           The systematic velocity of the approaching gas B component is 
                          about $-240$\,\kms\ with a 10\% uncertainty, while for the receding gas component, 
                          Rb+Rr, the overall systematic velocity is about $170$\,\kms\ with 15\% uncertainties.
	                      $^{(a)}$: The \htop\ line of G09v1.97 fitted here is dominated by 
	                      \htop(\t202111)$_{({5/2}\text{--}{3/2})}$ (rest-frame frequency at 
	                      742.1\,GHz). The contribution from the 742.3\,GHz line 
	                      \htop(\t211202)$_{({5/2}\text{--}{3/2})}$ is negligible.
	                      $^{(b)}$: The line \htop(\t211202)$_{({5/2}\text{--}{5/2})}$ 
	                      with rest-frame frequency of 746.5\,GHz is dominating the emission.
	                      The contribution from the 746.3\,GHz \htop(\t202111)$_{({3/2}\text{--}{3/2})}$ 
	                      can be neglected. Considering the similarity between the \hto\ 
	                      line and the \htop\ lines in the line profiles 
	                      \citep[][and this work]{2016A&A...595A..80Y}, the \htop\ line is expected 
	                      to have a similar profile structure. However, due to a gap between 
	                      the two spectral windows (indicated by green stripe) where the red component 
	                      of the line resides, the fitted values in the table is rather likely to be 
	                      close to that of the approaching gas component B.
	                      $^{(c)}$: H$_2^{18}$O(\t211202) is blended with N$_2$H$^+$(8--7), see text 
	                      for detailed discussions.
	                      $^{(d)}$: Because the line is weak and its blue-shifted part is partially
	                      in the spectral window gap, this component is rather representing the 
                          dominant receding gas Rb+Rr.  
}
   \label{table:obs-results}
\end{table*}
\normalsize

We note that the blue-shifted component (B) in the spectrum 
slightly alters the CO redshift from $z$\,=\,3.634, derived from the lower 
S/N IRAM\,30m data \citep{2017A&A...608A.144Y} where only the dominant 
red-shifted velocity component (consisting of the components Rb and Rr) was detected, 
to $z$\,=\,3.632. This new redshift is defined by the central line 
position as the center of the full width at zero intensity, representing 
the overall redshift of the entire system. We will consistently use 
$z$\,=\,3.632 as the redshift for G09v1.97 in this work.

The spectra of the \co65, \htot211202, and \htop\ lines have very similar 
profiles (Fig.\,\ref{fig:img:g09-spec}), composed of three Gaussian 
components, ``B'', ``Rb'', and ``Rr''. The observed velocity-integrated flux ratios 
between the three components are similar for the three lines within the 
uncertainties. The overall profiles display pronounced asymmetries with a 
strong red-shifted peak and a weak blue-shifted wing. The blue-shifted 
wing shows a single approaching gas component B centered at velocity of 
$-240$\,\kms, with a \FWHM\ of $\sim$\,300\,\kms\ and most of its fluxes 
resides at negative velocities. The strong (receding) red-shifted peak 
emission feature can be explained by the two Gaussian components Rb and Rr, 
dominating the fluxes in positive velocity channel bins. The linewidths 
for Rb and Rr are somewhat different (see Table\,\ref{table:obs-results}), 
and both components are close in velocity, peaking at 100 and 237\,\kms, 
respectively. This suggests that Rb and Rr are likely closely related.
The peak flux ratio of Rr to Rb is $\sim$\,1.3 for the lines. The possible 
origin of the asymmetrical line profile could be an intrinsic asymmetrical 
line profile or/and differential lensing. The peak of the B component 
is $\sim$\,4 times weaker compared to the overall receding gas (Rb+Rr), 
and the linewidth is 1.5 times narrower (Table\,\ref{table:obs-results}). 
The fact that the velocity separation between the component B and the group 
Rb/Rr is much larger than the velocity separation between Rb and 
Rr is an indication that B is likely to be a distinct velocity component. 
We will further discuss the nature of these velocity components in the 
following sections using position-velocity (PV) diagrams in the source 
plane after correcting for the gravitational lensing. 

The significant similarity among the gas tracers strongly suggests that 
the emitting regions overlap 
for the \co65, \htot211202, and \htop\ lines, which is also supported by 
their similar moment maps (Fig\,\ref{h2o:g09-mom-image}). All these similarities 
in the spatial and kinematical distributions for the lines indicate that 
these gas tracers and the dust continuum emission are closely related to 
the similar active star-forming regions \citep[see][who analyzed the gas 
excitation reaching a similar conclusion]{2016A&A...595A..80Y, 2017A&A...608A.144Y}.
We will further discuss the possible scenarios of the asymmetrical line 
profiles and compare the spatial and kinematical structure of \co65\ and 
\htot211202\ in the source plane in Section\,\ref{section:source_plane}. 

We derive the apparent line luminosities, $\mu L_\mathrm{line}$ and 
$\mu L^{\prime}_\mathrm{line}$, via \citep[see][]{1992ApJ...387L..55S} 
\small
\begin{align*}
\label{eq:luminosity}
\begin{split}
L_\mathrm{line}= 1.04 \times 10^{-3} 
               ({I_\mathrm{line} \over \mathrm{Jy\,km\,s^{-1}}}) 
               ({\nu_\mathrm{rest} \over \mathrm{GHz}}) 
               (1+z)^{-1}
               ({D_\mathrm{L} \over \mathrm{Mpc}})^2 \,L_\odot,\hspace{50pt}\\
L^{\prime}_\mathrm{line}= 3.25 \times 10^{7} 
               ({I_\mathrm{line} \over \mathrm{Jy\,km\,s^{-1}}}) 
               ({\nu_\mathrm{obs} \over \mathrm{GHz}})^{-2} 
               (1+z)^{-3}
               ({D_\mathrm{L} \over \mathrm{Mpc}})^2 \,\kkmspc,
\end{split}
\end{align*}
\normalsize
from the observed line flux densities. $I_\mathrm{line}$ is the velocity
integrated line flux, $\nu_\mathrm{rest}$ and $\nu_\mathrm{obs}$
are the rest-frame and observed frequencies, and $D_\mathrm{L}$ is
the luminosity distance. The apparent CO line luminosity is around
(4--10)$\times10^{8}$\,\lsun\ or (4--9)$\times10^{10}$\,\kkmspc, which is
about $7\times10^{-6}$ weaker than the apparent \lir, while for the \hto\ 
line, the line luminosity is about 2--3 times lower than the 
CO line (for B, Rb, and Rr), i.e., about (2--4)$\times10^{8}$\,\lsun\ 
or (1--3)$\times10^{10}$\,\kkmspc. The \htop\ lines are $\sim$\,3--4 times 
weaker than the \htot211202\ line (Table\,\ref{table:obs-results}). 

At a rest-frame frequency of $\sim$\,745\,GHz, we observe a 
5-$\sigma$ emission line with an integrated flux of $0.5\pm0.1$\,Jy\,\kms\ 
and a linewidth of $250\pm58$\,\kms. A similar emission line has also been 
observed at $\sim$\,745.3\,GHz in another {\it H}-ATLAS source, NCv1.143 
({\it H}-ATLAS\,J125632.7+233625), at a significance of 3-$\sigma$ \citep{2016A&A...595A..80Y}, 
and was tentatively identified as the H$_2^{18}$O(\t211202) line.
The detected 5-$\sigma$ emission at $\sim$\,745\,GHz corresponds to the position
of the Rb+Rr component of the rest-frame 745.320\,GHz H$_2^{18}$O(\t211202) line.
Therefore, the line is likely to be H$_2^{18}$O(\t211202).
Strong absorption lines of H$_2^{18}$O which are excited by far-IR pumping 
have been observed in local luminous IR galaxies (ULIRGS, defined by 
10$^{12}$\,\lsun\,<\,\lir\,<\,10$^{13}$\,\lsun) with {\it Herschel}-PACS, 
indicate enhanced abundance ratios of H$_2^{16}$O/H$_2^{18}$O $\sim$\,70 
(e.g., in Arp\,220, \citealt{2012A&A...541A...4G}, and, with possibly higher 
ratios in Mrk\,231, \citealt{2010A&A...518L..41F, 2014A&A...561A..27G, 
2018ApJ...857...66G}) suggesting that this line might be 
ubiquitous in starburst galaxies, and could potentially help us constrain 
the abundance of H$_2^{18}$O at high redshift.

Another possible identification of this emission line is 
N$_2$H$^+$(8--7) at 745.210\,GHz. However, we rule out this identification 
for the following reasons. N$_2$H$^+$ is known to be a tracer of the 
quiescent gas associated with dense, cold star-forming cores 
\citep[e.g.,][]{1996ARA&A..34..111B, 2002ApJ...572..238C} and is
therefore less likely to be detected in a intense starburst like  
G09v1.97. The first high-redshift detection of N$_2$H$^+$ was discussed  
in \citet{1996Natur.379..139W}, but only absorption against the background  
radio source PKS~1830$-$211. Recently, \cite{2017A&A...608A..30F} 
reported an emission line at 94.83\,GHz in the quasar APM\,08279+5255, 
which they tentatively identified as N$_2$H$^+$(5--4) without completely 
ruling out other possibilities, including the detection of a low-$J$ CO 
line from the foreground deflector. \cite{2015A&A...579A.101A} derived a 
ratio of $\approx$\,2 between the ground transition lines of HCO$^+$ and N$_2$H$^+$ in
nearby active galaxies. If this value is also valid for higher energy 
levels, we would expect a similar ratio for HCO$^+$(8--7)/N$_2$H$^+$(8--7). 
The observed integrated flux of HCO$^+$(5--4) in G09v1.97 is about 
0.5\,Jy\,\kms\ \citep{2017PhDT........21Y}. Assuming that the flux 
ratio of HCO$^+$(8--7) to HCO$^+$(5--4) is $\sim$0.7 
\citep{2017ApJ...849...29I}, the expected flux of N$_2$H$^+$(8--7) 
in G09v1.97 would be $\sim$\,0.2\,Jy\,\kms, which is less than 
half of the currently measured flux.

Based on the above arguments, we conclude that the emission feature 
detected at the rest-frame frequency of $\sim$\,745.32\,GHz in G09v1.97 
is most likely to be the H$_2^{18}$O(\t211202) line. 
This line is only seen in the red-shifted velocity 
component (dominated by Rr) and not in the weaker B component, 
which we attribute to component B having lower S/N.

\section{Lens Modeling}
\label{section:lens_model}

In order to derive the intrinsic properties of G09v1.97, a lens model 
needs to be built using both the high-resolution ALMA imaging data and 
the optical/near-IR images to constrain the gravitational potential 
of the deflectors. We stress that the main focus of this 
paper is to study the properties of the background lensed source, and 
the detailed structure of the deflecting mass distribution will be 
presented in a subsequent study.

\begin{table*}[!htbp]
\centering
 \small
 \caption{Lens modeling results.}\label{tab:lensmodel-G09}
 \setlength{\tabcolsep}{1.35em}
 \begin{tabular}{rccccc} 
 \toprule 
 \multicolumn{6}{c}{Parameters of SIE mass components}                                                                                \\[.2cm]
                           & $x_{\rm def}$            & $y_{\rm def}$             & $q_{\rm def}$    & $PA_{\rm def}$ & $\sigma_v$    \\
                           &  (arcsec)                & (arcsec)                  &                  & (deg)          &   ($\kms$)    \\ 
\midrule
Southeast $z=0.626$ lens (G1)  &\llap{$-$}$0.22 \pm 0.01$ &\llap{$-$}$0.12 \pm 0.01 $ & $0.53 \pm 0.03 $ & $105 \pm 3 $  & $ 142 \pm 2 $ \\
Northwest $z=1.002$ lens (G2)  &          $0.12 \pm 0.01$ &          $0.37 \pm 0.01 $ & $0.82 \pm 0.03 $ &  $12 \pm 4 $  & $ 181 \pm 2 $ \\ 
\bottomrule
 \end{tabular}

\smallskip

\setlength{\tabcolsep}{0.592em}
\begin{tabular}{rccccccccc}
\toprule 
\multicolumn{10}{c}{Parameters of reconstructed source }                                                                                                                                                         \\[.2cm]
                    &$x_{\rm src}$                   & $y_{\rm src}$         & $q_{\rm src}$    & $PA_{\rm src}$   & $R_{\rm eff}$      & $S_{0.87}$ & $S_{1.94}$  & $S_{\rm 0.87}/S_{\rm 1.94}$  &$\mu$                   \\
                    &       (arcsec)                 & (arcsec)              &                  & (deg)            &   (arcsec)         &   (mJy)    &   (mJy)     &                              &                        \\
\midrule                                                                                                                                                                                    
Compact  continuum  &    $  0.02 \pm 0.01 $          &  $  0.02 \pm 0.01 $   & $0.31 \pm 0.03 $ &  $9 \pm  3$      & $0.09 \pm 0.01$    &    {${3.2\pm0.5}$}    &  {$0.59\pm0.08$}       &  $\,\,5.4 \pm 0.4$           & $ 10.2\pm 0.9 $        \\ 
Extended continuum  &    \llap{$-$}$0.01 \pm 0.01 $  &  $  0.04 \pm 0.01 $   & $0.82 \pm 0.05 $ &  $29 \pm  16$    & $0.19 \pm 0.01$    &    {${7.0\pm1.4}$}    &  {$0.27\pm0.08$}      &  $26^{+9}_{-6}$              & $ 10.5^{+0.6}_{-0.5} $  \\ 
\bottomrule
\end{tabular}
   \begin{tablenotes}[flushleft]
   \small
	 \item\textbf{Note:} 
           Positions are expressed in arcseconds relative to the central coordinates 
           (J2000, RA 08:30:51.156, Dec $+$01:32:24.35). The position angles of the major 
		   axis of ellipses are defined in degrees east of north. The axis ratio is the one 
           of the mass and not of the potential. Values with subscription 
           ``def'' are for deflectors G1 and G2 while ``src'' are the ones for the background source.
           $F_{\rm 0.87}/F_{\rm 1.94}$ is the 0.87\,mm-to-1.94\,mm 
           dust continuum flux ratio integrated over the entire source.
   \end{tablenotes}
\label{table:lens-mod}
\end{table*}
\normalsize

Similar to the parametric lens model in \citet{2013ApJ...779...25B}, 
the two foreground deflectors (G1 and G2, see Fig.\,\ref{fig:g09-image}) 
are assumed to be singular isothermal ellipsoid (SIE) mass distributions 
centered on the two foreground galaxies. The parameter encoding the 
strength of the deflector, namely the depth of the lensing potential, 
is represented by the velocity dispersion $\sigma_v$ (note here $\sigma_v$ 
is not the velocity dispersion of the ISM gas). The velocity dispersion 
of the lens model is related to the Einstein radius by 
$R_{\rm Ein}= 4 \pi \left( \sigma_v/c\right)^2\,(D_{\rm ds}/D_{\rm d})$, 
in which $D_{\rm d}$ and $D_{\rm ds}$ are the distance between the 
observer and deflector, and the distance between the deflector and the 
lensed object, respectively. The minor-to-major axis ratio $q$ and 
orientation of the major axis in the plane of the sky (position angle,
$\mathit{PA}$) are left free and explored over the parameter space.

The lens model is based on the {\tt sl\_fit} lens inversion code, 
following the method described in \citet{2011ApJ...738..125G}. 
Here we adopted a Markov chain Monte Carlo (MCMC) method, implementing 
the standard Metropolis-Hastings algorithm. It explores the space
of lens model parameters and builds samples of the posterior probability 
distribution function. The {\tt sl\_fit} code is mostly tailored to 
fit optical/near-IR data. Nevertheless, we are able to account for 
synthesized beam and noise correlation in the \texttt{CLEAN}ed images 
for our ALMA data using the methods described in \citet{2011ApJ...738..125G}. 
Although fitting visibilities in the uv-plane would overcome the 
caveats related to side-lobes and correlated noise of interferometric 
data \citep[e.g.,][]{2012ApJ...756..134B, 2013ApJ...779...25B, 
2013ApJ...767..132H, 2015MNRAS.452.2940N, 2016ApJ...826..112S, 
2017ApJ...836..180L, 2018MNRAS.475.3467E}, this is not implemented 
yet in {\tt sl\_fit}. \citet{2018MNRAS.476.4383D} showed that 
image- and uv-plane model fitting can yield highly consistent 
results with ALMA data with sufficiently high uv-plane coverage and thus 
small dirty beam side-lobes. Since our ALMA data has such sufficiently  
high uv-plane coverage and S/N, we safely assume that any information 
loss in the source plane as a result of our image-plane analysis 
is negligible. We leave a thorough comparison of visibility versus 
image-space fitting for future studies using higher angular resolution 
ALMA images. However, even with higher resolution, we do not expect any 
significant changes.

\subsection{Fitting the dust continuum emission and the mass model}
\label{section:lensmodel1}

\begin{figure*}[htbp]
\includegraphics[width=0.992\textwidth]{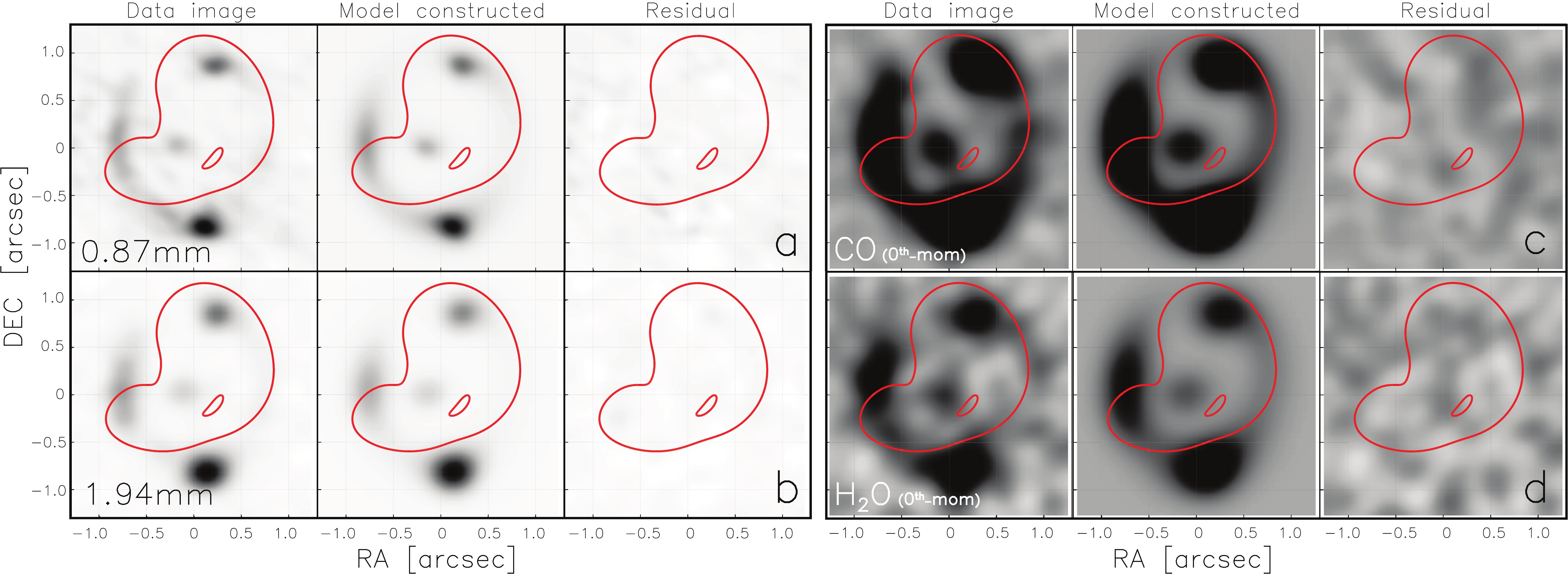}
\vspace{-0.2cm}
\caption{
{\it Panels a-c}: Image-plane lens modeling results of the 
1.94 and 0.87\,mm dust continuum, \co65\ and \htot211202\ 
line emission for G09v1.97. The panels are grouped in 
horizontal sets of three sub-panels each showing, from left to 
right: i) observed image; ii) reconstructed image from the lens 
model; and iii) residuals from the difference between observed 
and model images. The red lines represent the critical curves. 
The central coordinates are given in Table.\,\ref{table:lens-mod}. 
The figure clearly demonstrates that the lens model can recover 
the fluxes of the dust continuum and line emission accurately. 
The corresponding source-plane images are show in Fig\,\ref{fig:source-line}. 
All the residuals are well within $\pm$2.5\,$\sigma$, showing 
that all the modeled images of the dust continuum and the emission 
lines agree very well with the overall flux distribution.
}
\label{fig:resid-g97}
\end{figure*}

Our lens model places the deflectors at their corresponding spectroscopic 
redshifts ($z$\,=\,0.626 for G1 and $z$\,=\,1.002 for G2). Following the 
prescription of \citet{slacs6}, we allow for the slight lensing of G2 by 
G1 via a flat prior on the position of G2 and we check that the image-plane 
location of its center after modeling coincides with that observed. 
For G1, we apply a Gaussian prior of width $0\farcs1$ centered on its observed 
position to accommodate absolute astrometric uncertainties. We have also 
run models where both deflectors are placed at the same redshift of $z$\,=\,0.626, 
finding very similar results apart from the central image which becomes 
slightly more magnified.

Because the dust continuum images have the best S/N values, we used the 1.94 
and 0.87\,mm dust continuum images (Fig.\,\ref{fig:g09-image}) to constrain 
simultaneously the mass distribution (SIE model) of G1 and G2 and the 
dust emission from the background source. The resulting best-fit mass 
model is then used to reconstruct, in the source plane, the line emission 
in each velocity channel. This approach is found to capture most of the 
information content of the data. Deeper and higher resolution data would 
enable us to perform a joint fit to the continuum and line emission, 
and eventually, provide the greater flexibility of creating a 
``pixelated source'', as was the case for the high-resolution 
observations of SDP\,81 \citep[e.g.,][]{2015MNRAS.452.2258D}.

\begin{figure*}[!htbp]
\includegraphics[width=1.0\textwidth]{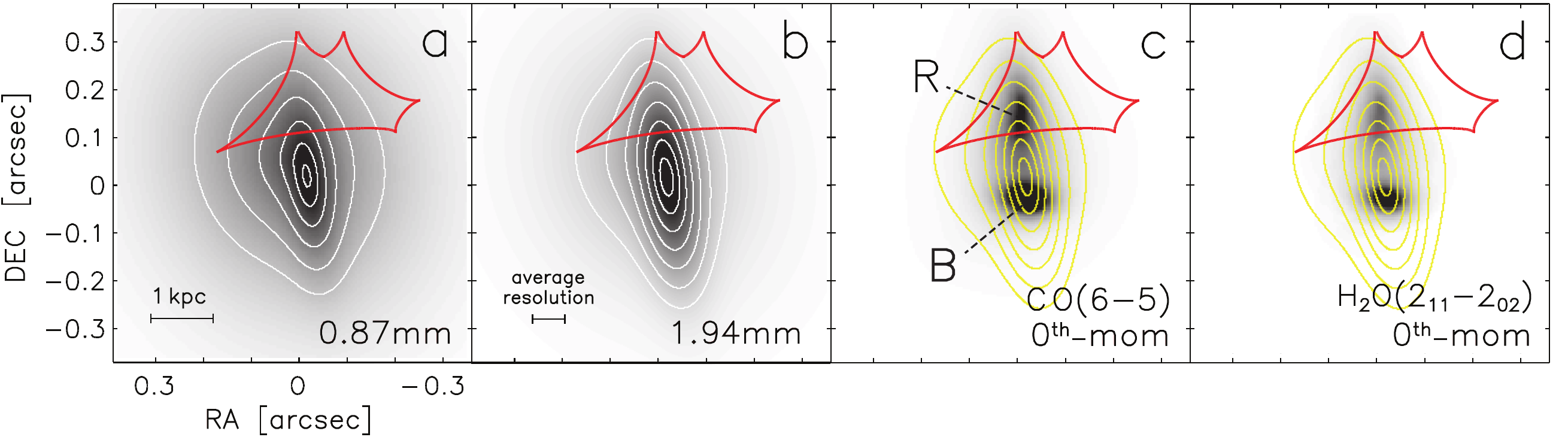}
\vspace{-0.6cm}
\caption{
Reconstructed images of the dust continuum and line emission of 
G09v1.97 using the lens model, displaying: ({\it a, b}) the 
images and contours showing the source-plane 0.87 and 1.94\,mm 
dust continuum emission, and ({\it c, d}) the velocity-integrated 
\co65\ and \htot211202\ line emission with yellow contours 
of the 1.94\,mm dust emission overlaid for comparison. The 
central coordinates are the same as in Fig.\,\ref{fig:resid-g97}. 
The caustic lines are shown in red. The flux ratio of the 
two dust continuum disk components at 0.87 and 1.94\,mm are 
different, indicating that the two components have 
either different dust temperatures and/or different submm optical 
depths. The average resolution indicated in panel {\it b} is 
derived from $0\farcs2/\!\sqrt{\mu}$\,=0\farcs06 (0.4\,kpc). 
The gas emission traced by \co65\ and \htot211202\ is 
configured in two distinct components separated by 1.3\,kpc in projection. 
The R component is associated with the red-shifted 
velocity bins of the line spectra, namely Rr and Rb of the 
line profile, while the B component corresponds to the blue-shifted 
part of the line profile as shown in Fig.\,\ref{fig:img:g09-spec}. The dust 
continuum (rest-frame 188 and 419\,$\mu$m) peaks between 
the two gas components and a bit towards the south. The gas 
emission is more extended than the dust continuum (comparing 
the half-light radii, see Section\,\ref{section:source_plane:image} 
for details) consistent with the analysis of the 
image plane.
}
\label{fig:source-line}
\end{figure*}

As discussed in \citet{2013ApJ...779...25B}, a single S\'ersic 
component of index $n\simeq1.8$ was found to provide a good fit to the 
lower-spatial-resolution SMA data. With the improved depth and spatial 
resolution of our ALMA data, we can perform modeling with a 
more complex intrinsic light distribution. Therefore, we assume that 
the dust emission consists of two exponential profiles with several free 
parameters, e.g., positions ($x$, $y$), sizes (effective 
radius $R_\mathrm{eff}$), ellipticities ($q$, which equals the 
minor-to-major axis ratio) and orientations ($\mathit{PA}$). Those 
parameters can differ between each source component, although, by 
construction, the geometry (size, orientation, position) remains the 
same at 1.94 and 0.87\,mm. Only the flux can differ 
across the two bands. In other words, each of the two background 
components has a constant magnification across its entire far-IR 
spectral energy distribution (SED). Two such sources sharing the 
same center would mimic a unique source with an SED gradient, should 
the data demand it, but these model assumptions also allow us to 
model two spatially distinct internally homogeneous sources.
The Bayesian approach used in {\tt sl\_fit} requires a clear definition 
of modeling priors. For the lensing potential, we assume flat priors 
for the position of G2 and a Normal prior for the width $0\farcs1$ centered 
on the observed position of the G1 galaxy. This is to take into account the deflection 
of the more off-axis galaxy G2 by source G1 which lies closer to the main axis of the
deflection. The orientation of 
the major axis has a uniform prior whereas both axis ratios have a 
Normal prior centered on 0.5 and of width 0.2 and set to zero outside the 
range $[0,1]$. 
The prior on both Einstein radii is uniform in the 
range $[0\farcs1,1\farcs1]$. For the exponential profiles used for the 
morphology of  
the source plane dust emission, we apply a uniform prior in the range 
$[-0\farcs2,0\farcs2]$ on both coordinates of both sources\footnote{This 
relatively narrow window was guided by previous models exploring a wider 
space of priors. Restricting the range simply helps to speed up the convergence of the model.}. Both axis ratios 
have a uniform prior distribution in the range $[0.1,0.9]$ and orientations 
are uniform on the circle. Intrinsic source plane fluxes in both frequency 
channels are uniformly bound between 0 and a conservative upper limit set 
at 1.5 times the total image plane flux in the corresponding 
velocity channel. Finally, we applied a Normal prior for the effective radius 
centered on $0\farcs1$ and of width $0\farcs03$ multiplied by a sharp 
uniform prior in the range $[0,0\farcs2]$.

The results of the lens model and the reconstructed image-plane 
images at 1.94 and 0.87\,mm are shown in panel {\it a} and {\it b} 
of Fig.\,\ref{fig:resid-g97}. The constraints (median and $68\%$ 
confidence level intervals on the marginal distributions) of the 
model parameters defining the mass model and the source-plane dust 
continuum emission are provided in Table\,\ref{tab:lensmodel-G09}. 
For a few relevant parameters, marginal posterior distributions and 
pair-wise scatter plots are shown in Fig.\,\ref{fig:cornerplot} in 
Appendix\,\ref{appendix:models} to illustrate possible parameters 
degeneracies. Despite its apparent simplicity, the model is able 
to reproduce most of the light distribution, leaving almost all 
residuals close to the noise level (<\,2.5\,$\sigma$). One can 
recognize a nearly fold-like configuration, with the faint parts 
of the source straddling the caustic. The critical lines resulting 
from the best fit mass distribution clearly reflect the bimodality 
of the foreground mass. The double nature of the deflecting system 
(G1 and G2) introduces a central de-magnified image that is observed 
and well reproduced by the model.

The two deflectors have relatively low masses with an Einstein radius 
of $0\farcs39 ^{+0.01}_{-0.01}$ for G1 at $z$\,=\,0.626  
and $0\farcs63^{+0.01}_{-0.01}$ for the more distant G2 at 
$z$\,=\,1.002, corresponding to $\sigma_v=$\,$142 \pm 2$\,\kms\ and 
$\sigma_v=$\,$181 \pm 2$\,\kms\, respectively. The two foreground 
galaxies have different shapes, with the ellipticity of G2 appearing 
to be well aligned with that of the host galaxy, and an orientation 
that is consistent with the shear generated by G1.

The source-plane dust continuum images at 1.94 and 0.87\,mm reconstructed
by the lens model are displayed in Fig.\,\ref{fig:source-line}. 
The model requires two nearly concentric dust components with 
a very small separation, $0\farcs04 \pm 0\farcs01$. One   
component is compact with a half-light radius, $R_{\rm eff,\,1}$\,=\,$0.63 \pm 0.15\, {\rm kpc}$, 
has a prominent north-south elongation, and contains the 
peak of the surface brightness distribution (the ``core''). There is an additional 
extended envelope which is more circular and substantially larger with 
$R_{\rm eff,\,2}$\,=\,$1.37 \pm 0.05\, {\rm kpc}$. At 0.87\,mm, the 
compact source is $0.46^{+0.09}_{-0.08}$ times brighter than the 
extended component, whereas this ratio rises to $2.2^{+1.0}_{-0.6}$ 
at 1.94\,mm, suggesting that either the dust temperature and/or 
the submm optical depths might be different in the core and  
envelope (see Table\,\ref{tab:lensmodel-G09}). The compact 
core and the envelope experience a similar overall magnification. 
The total magnification is of order $\mu_{\rm tot}$\,=\,$10.3\pm 0.5$, 
somewhat higher than the value $\mu_{\rm tot,\,B13}$\,=\,$6.9\pm0.6$ 
derived by \citet{2013ApJ...779...25B}.

One should note that, although our double-disk model captures most of 
the dust continuum flux and the overall structure of the source, 
the 0\farcs2--0\farcs3 resolution ALMA continuum images could 
potentially even capture the flux variations at smaller scales.  
The average scale magnification can be inferred with $\sqrt{\mu}$\,$\sim$\,3.2.
With such a magnification, the 0\farcs2--0\farcs3 continuum image 
will be resolved into average scales of $\sim$0.4\,kpc (0\farcs06).
This has been further tested with  
\texttt{PyAutoLens}\footnote{\url{https://github.com/Jammy2211/PyAutoLens}} 
\citep{2018MNRAS.478.4738N}, which reconstructs the source-galaxy 
using an Adaptive Voronoi tesselation as opposed to analytic S\'ersic 
light profiles. The analysis converges to the same lens model and 
reconstructs a source galaxy with the same global structure as the 
double-disk model, yet it reveals subtle variations on smaller scales 
comparable with the averaged magnified scale of angular resolution, 
or even slightly smaller at locations close to the caustic. Nevertheless, 
the discussion on such variation structures at scales $<$\,0.4\,kpc 
are beyond the scope of this paper. This lens modeling therefore 
verifies our parametric lens model which we will use hereafter to 
determine the properties of G09v1.97.

\subsection{Fitting the CO and \hto\ Line Emission}
\label{section:lensmodel2}

The SIE parameters of the best (lens) mass model derived from the two dust 
continuum images were used as input to model the data cubes of the \co65\ 
and \htot211202\ emission lines. Due to the limited S/N at the edges 
of the spectra, we only performed such line-emission reconstructions 
in the source plane using the channel bins located in the velocity ranges 
within $\sim$\,$\pm$\,450\,\kms, which covers all the full widths at zero 
intensity. Performing the inversion channel per channel, 
we study the intrinsic source-plane line emission and, in particular, 
the spatial variations of the line of sight velocity distribution (LOSVD), 
similar to e.g., \citealt{2008ApJ...686..851R, 2011ApJ...742...11S, 
2015ApJ...811..124S, 2017ApJ...836..180L}. We focus on the integrated 
line emission map ($\rm 0^{th}$ moment of the LOSVD) and velocity field 
($\rm 1^{st}$ moment of the LOSVD).

In order to perform the inversion, the data cube was binned into 10
velocity channel bins of 105\,\kms\ from $-$478 to 467\,km\,s$^{-1}$ 
for the \co65\ and \htot211202\ line cubes. Given the relative simplicity 
of the continuum emission, and in order to obtain the simplest possible 
source model, we assume that, up to a normalizing flux constant, the 
emission is identical between the \co65\ and \htot211202\ lines, and 
each emission component is well represented by a single elliptical exponential profile whose parameters 
(center, ellipticity, orientation, effective radius, and fluxes) are 
estimated for each independent slice. This procedure is supported by the 
pronounced similarities in the spatial and the velocity structure of 
\co65\ and \htot211202\ (Fig.\,\ref{fig:mom1_ratio}). We thus fit for 
the 10$\times$7 parameters defining the source emission while using 
the foreground mass distribution derived from the dust continuum 
modeling (Section\,\ref{section:lensmodel1}). As with the modeling of 
this latter component, we apply the same priors on all of the source parameters. 
Hence, we obtain a model-predicted data cube in both the image and 
source planes. 
 
\begin{figure}[!htbp]
\centering
\includegraphics[width=0.465\textwidth]{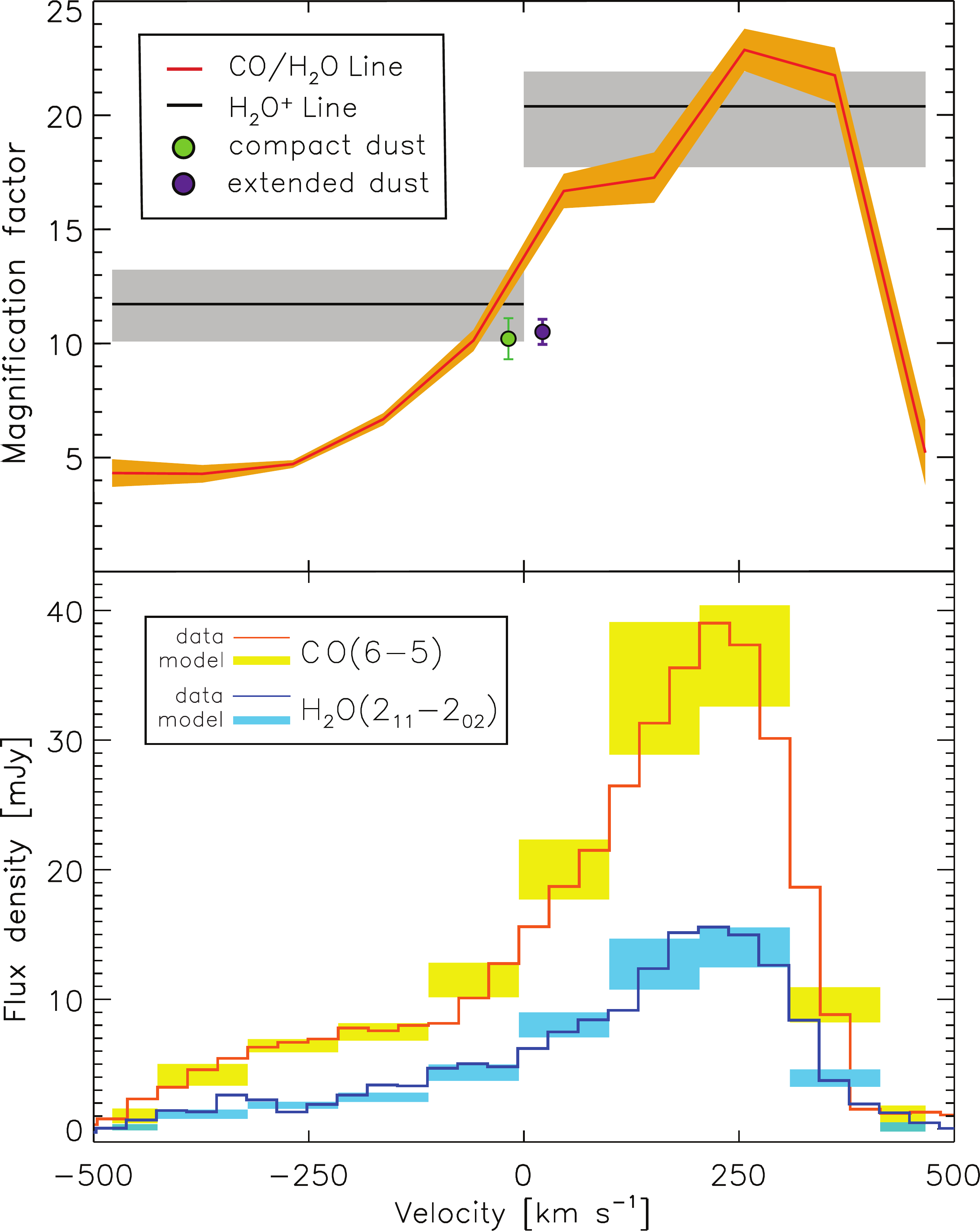}
\vspace{-0.2cm}
\caption{  
{\it Top panel}: Magnification factor as a 
function of velocity for the emission lines of CO, \hto,  
and \htop, and indicated for the two dust components (the 
legend at the top left corner indicates the line 
and point style for each component of emission).
{\it Bottom panel}: 
A comparison between the integrated spectrum of the observations  
and lens-model reconstructed data seen in the image plane. 
The filled regions show the model reconstructed using the bins in the image-plane 
spectra, while the histogram lines show the observed image-plane 
spectra (see the legend at the top left). The model overall predicts very well the spectra in the 
source plane.}
\label{fig:FluxCompar}
\end{figure}

As shown in panels {\it c} and {\it d} of Fig.\,\ref{fig:resid-g97}, 
it is clear that the lens model reproduces the overall fluxes of both the 
\co65\ and \htot211202\ line emission. Both the $\rm 0^{th}$ moment maps 
are well reproduced with our lens models and the residual shows no 
significant disagreement. As shown in Fig.\,\ref{fig:allvelchannels}, the 
model provides a good reconstruction of the line emission in each channel 
bin and therefore reproduces the entire velocity field. Furthermore, 
Fig.\,\ref{fig:FluxCompar} shows that the model reproduces the observed 
integrated line profiles, together with the magnification factor at each 
channel for the lines and the two dust components. 

We also explored the nature of the \htop\ emission, which has lower 
S/N values than either \co65\ or \htot211202. We performed a similar 
analysis for the \htop\ data cube by dividing the emission into two 
channels, and the results are shown in Fig.\,\ref{fig:FluxCompar} and 
Fig.\,\ref{fig:allvelchannels}.

We conclude that for all the lines, the image plane reconstruction is 
in good agreement with the observed imaging data, suggesting that our 
simple modeling assumptions already capture most of the information 
content of the current data. Increasing the complexity of the gas 
distribution would require more flexible modeling techniques (and more 
free parameters) which are not required with these data. Based on the lens 
inversion results, we will describe the intrinsic physical properties 
of the source including its kinematics and morphology in the source 
plane in the following sections.

\section{G09v1.97 in the source plane}
\label{section:source_plane}

\subsection{Morphology of the dust continuum and line emission}
\label{section:source_plane:image}

As shown in Fig.\,\ref{fig:source-line}, the molecular line 
maps of \co65\ and \htot211202\ have an overall distribution  
elongated along the north-south direction. The orientation of 
the line emission is 
similar to the dust continuum at rest-frame 
188 and 419\,$\mu$m, albeit more complex with indications of a 
bimodal structure. The source-plane dust continuum image  
shows a predominately elongated disk-like smooth distribution 
along the north-south direction, and which peaks between the two gas 
components, albeit slightly towards the southern one.

\begin{figure}[htbp]
\includegraphics[width=0.493\textwidth]{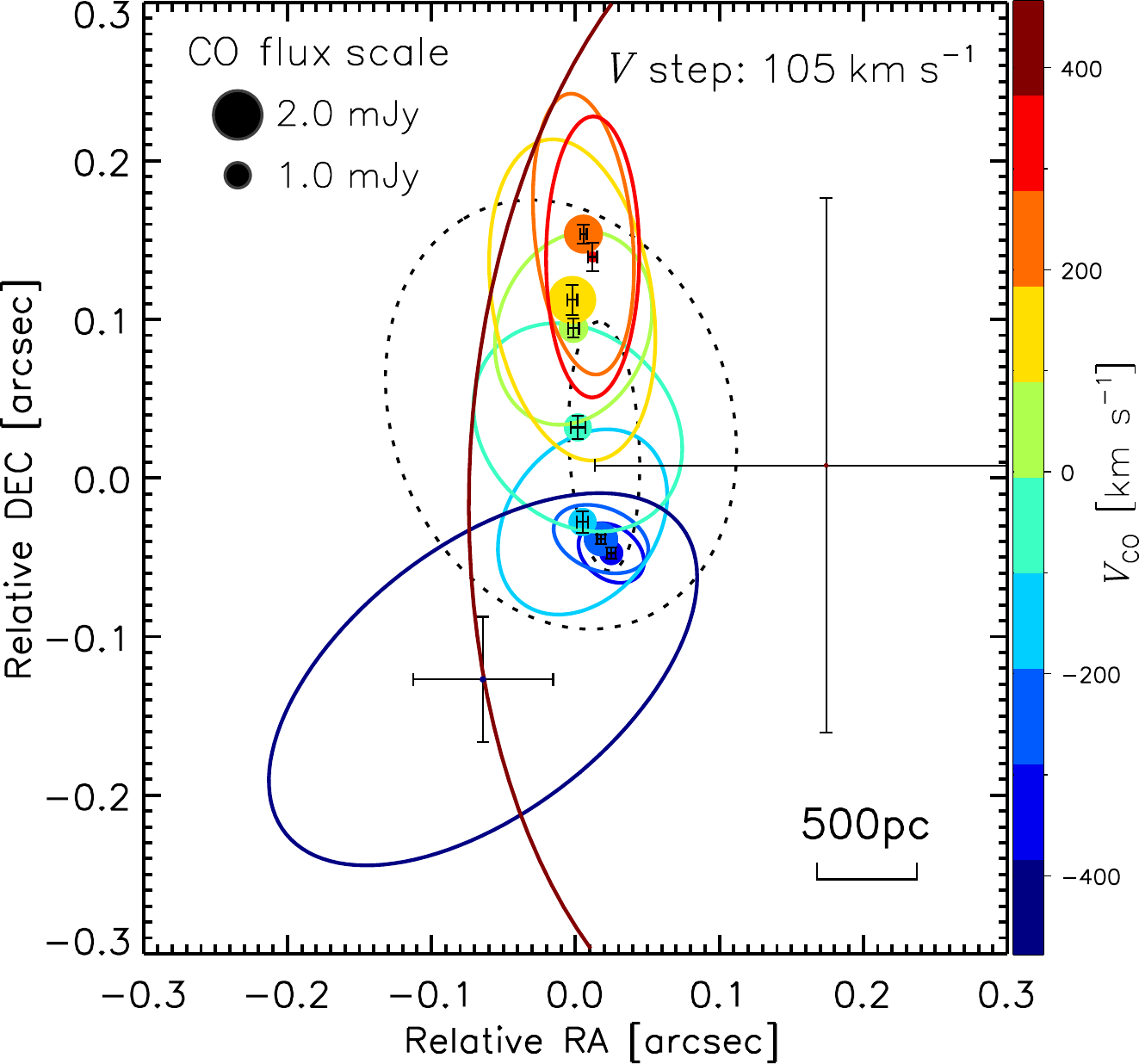} 
\vspace{-0.5cm}
\caption{
Ellipses showing the disk model for the \co65\ line 
emission per velocity channel bin (the central velocity of 
each disk is indicated by its color as indicated in the 
color bar). The parameters used to draw each ellipse in 
the diagram, i.e., half-light radius, $PA$, and ellipticity, 
are their median values. The points with error bars indicate 
the central positions and their uncertainties for each disk. 
The sizes and positions of the filled circles indicate 
the total flux and the center of the disk at each velocity 
bin (the flux scales are shown in the legend in the top 
left corner and the velocities are given by the color bar). 
The velocity bins are in steps of 105\,\kms\ (see text). 
The figure shows clearly a concentration of fluxes with 
approaching velocities at radii smaller than 
$\sim$\,0\farcs05 (0.4\,kpc), while the fluxes associated 
with the receding velocities show a significant gradual 
change in position and a larger overall radius of 
$\sim$\,0\farcs16 (1.2\,kpc).
}
\label{fig:centroids}
\end{figure}

To better demonstrate the spatial distribution of the gas emission 
and its velocity structure, Fig.\,\ref{fig:centroids} displays the  
gas emission per velocity channel bin as described 
in Section\,\ref{section:lensmodel2}. As clearly seen in the figure, 
the two distinct components in Fig.\,\ref{fig:source-line} of the 
0$^{\rm th}$ moment map of the gas emission have significantly  
different velocities. The northern disk is dominated by the emission 
from the red-shifted gas components (corresponds to ``R'' in 
Fig.\,\ref{fig:source-line}) while the southern component is dominated 
by blue-shifted gas (corresponds to ``B'' in Fig.\,\ref{fig:source-line}).  
There is also a gas bridge peaking at $\approx$\,$-$59\,\kms, which is 
located near the peak of the dust continuum components (dashed 
ellipse). The reader should note that the R component in the 0$^{\rm th}$ 
moment map corresponds to the ``Rb'' and ``Rr'' parts of the line profile 
observed in the image plane, while the B component is associated 
with the ``B'' part of the image-plane line profile (Fig.\,\ref{fig:img:g09-spec}). 
The gas components associated with red-shifted velocities are 
located within the caustic pattern and hence experience a 
stronger magnification ($\mu$\,$\sim$\,12) than the southern blue-shifted component 
($\mu$\,$\sim$\,5). This differential magnification contributes to 
the asymmetry of the image-plane integrated line profile, but only 
partially, because the intrinsic line-profile is asymmetric  
(Fig.\,\ref{fig:source-plane-spec}). Most of the flux from the gas emission 
originates in the two blobs, i.e., R and B, with only a minor 
contribution from gas between the two. For the southern B component, 
the centroids at different velocities bins (central velocity of each 
bin from $-$478 to $-$163\,\kms) are varying in a very small spatial 
range of 0\farcs03, neglecting the most blue-shifted velocity bin of 
$-$478\,\kms\ which has very low flux and large uncertainties. On the 
other hand, the northern R component shows a clear gradual variation 
over 0\farcs06 of the centroid positions from 47 to 362\,\kms (here we 
also neglect the 467\,\kms channel because of its low flux and high 
uncertainty). Such a difference suggests that R has a clear velocity 
shear, presumably rotation, while B is not kinematically 
resolved. We will discuss the kinematics in more detail in 
Section\,\ref{section:source_plane:kinematics}.

Fig.\,\ref{fig:centroids} also shows the half-light radius of the 
dust continuum and the line emission for each velocity channel bin, 
as indicated by the size of the ellipse. It shows that the overall 
size of gas is a bit more extended than the dust continuum emission 
(comparing half-light radii), with the extended part of the dust 
encompassing most of the line emission except for the most northern part  
of the R components. This is, to first order, consistent with 
the findings that the CO and \hto\ lines and submm dust continuum 
are presumably coming from similar active star-forming regions 
\citep{2016A&A...595A..80Y, 2017A&A...608A.144Y}. Yet, the detailed 
structure of the dust and gas have two different patterns, 
namely a smooth single disk-like structure for the dust continuum 
and a bimodal-disk distribution for gas.

\bigskip

The northern red-shifted gas R is somewhat more diffuse than the southern 
blue-shifted component B. Both gas components bracket the dust continuum 
emission. In order to better characterize the morphology of the molecular 
gas traced by \co65\ (and \htot211202), we performed a fit to 
the 0$^{\rm th}$ moment map of \co65\ in the source plane with a disk model 
using \texttt{IMFIT} \citep{2015ApJ...799..226E}. The central positions, 
position angles, ellipticities, S\'ersic indexes, and half-light radii are 
used as parameters to describe the disks in the model. One should note that 
the 0$^{\rm th}$ moment image is a result of the linear sum of the disk model 
fitted per velocity bin (Fig.\ref{fig:centroids} as described in 
Section\,\ref{section:lensmodel2}). The main purpose of fitting a disk model to 
the 0$^{\rm th}$ moment map is to estimate the overall size, ellipticity,  
and the position of the components seen in the map. Therefore, we used three 
disk-components to capture the features of the R and B components, 
also taking into account the gas which lies between the two components. The main 
parameters of this fitting are the half-light radius ($R_{\rm eff}$), 
central position, ellipticity ($e$), and position angle ($PA$). The parameter 
S\'ersic index used in the \texttt{IMFIT} is only for the propose of 
constraining the entire parameter set and it does not influence the final 
results in any significant way. We will also not discuss the gas between the 
two components because of its complex morphological structure and the fact that 
its parameters are less well constrained due to the relative low fluxes from 
this emission region. Using bootstrap 
resampling, the uncertainties from the fit are determined with uncertainties of 10\%. 
The best fitting model determined by minimizing $\chi^{2}$ (best-fit $\chi^{2}$\,=\,0.1) 
provides a good agreement with the overall morphology of the 0$^{\rm th}$ moment 
map, with insignificant residuals (maximum differences are less than 1\% of the emission). The 
best fit half-light radius, $R_{\rm eff}$ for R and B are 0\farcs10 (0.8\,kpc) 
and 0\farcs04 (0.3\,kpc), respectively. The corresponding ellipticities $e$ 
are 0.59 and 0.34, while the position angles are 4$^\circ$ and 97$^\circ$, 
for R and B, respectively. The two disks, R and B with semi-major 
axis half-length, $a_{\rm s}$ (define as \,$R_{\rm eff}/\sqrt{1-e}$) of 1.2\,kpc 
and 0.4\,kpc, are separated by a projected distance of 0\farcs18 (1.3\,kpc). 
Therefore, the overall size of the gas emission traced by \co65\ is $\sim$\,1.5 
times larger than the dust emission size (not necessarily the size of the 
dust distribution), which is consistent with the estimates found in the image plane. 
Assuming that both R and B are thin disks, the inclination angle can 
be derived from the minor to major axis ratio, $b/a$ (define as $1-e$), as 
$\cos^2{i}={((b/a)^2-q_0^2)/(1-q_0^2)}$, in which the value of $q_0$ indicates the 
intrinsic thickness of the disk \citep{1926ApJ....64..321H}. Choosing a typical value 
of $q_0=0.2$ for disk galaxies \citep[e.g.,][]{1946MeLuS.117....3H,1984AJ.....89..758H}, 
we infer inclinations for R and B of 82$^\circ$ and 66$^\circ$, respectively.

The projected separation between the northern 
and southern gas emission peaks is 1.3\,kpc with a difference in velocity 
of order $450$\,\kms and there is a significant difference in the spatial 
structure between the dust continuum and the molecular gas emission (the 
former shows a one-fold disk-like structure while the latter shows two 
blobs on both sides of the dust peak). This can either be explained 
by two distinct galaxies separated by a projected distance of 1.3\,kpc or
two (very luminous) dusty star-forming clumps within a single rotating 
disk with a size of 0.4--1.2\,kpc. Considering the first possible scenario, such 
a separation is consistent with the observations of local ULIRG mergers, 
which are in a close-to-coalesce phase or already coalescing 
\citep[e.g.,][]{1999AJ....118.2625R, 2000AJ....119..991S, 2015A&A...577A.119C}. 
Also the compact size of each merging galaxy is consistent with size estimates 
for other SMGs 
\citep[e.g.,][]{2008ApJ...680..246T, 2014ApJ...782...68T, 
2015ApJ...812...43B, 2015A&A...576A.127S, 2015ApJ...799...81S, 
2016ApJ...833..103H, 2016ApJ...826..112S, 2017ApJ...846..108C, 
2018ApJ...863...56C}. On the other hand, in the single-rotating-disk scenario, 
if the R and B components in G09v1.97 are within a single SMG, their 
relative large sizes compared to gas clumps ($\lesssim$\,100--300\,pc) 
derived from high-resolution observations in other SMGs 
\citep{2015ApJ...806L..17S}, suggest that they are unlikely to be the 
resolved individual star-forming clumps, but rather two collections of clumps. 
Additionally, we tested this single-SMG scenario by fitting a 
tilted-ring model \citep[e.g.,][]{1987PhDT.......199B} to the source-plane 
\co65\ data cube using \texttt{$^\mathrm{3D}$Barolo} 
\citep[a 3D-Based Analysis of Rotating Objects from Line Observations,][]{2015MNRAS.451.3021D}. 
We find that the rotating-disk model cannot explain the reconstructed 
source-plane data cube. The best-fitting model has a significant 
under-prediction for the red-shifted part of the line emission. 
Together with the spatial mismatch between the dust emission and the 
molecular line emission in the source plane (Fig.\,\ref{fig:source-line}), 
the poor fit likely rules out that the source is made of several clumps 
in a single rotating galaxy. Therefore, we conclude that the system is 
made of two compact merging galaxies with a small separation.

The ratios of 1.94 and 0.87\,mm dust continuum 
flux densities are different for the compact and extended components  
(Table\,\ref{tab:lensmodel-G09}). At the shorter wavelength, 
rest-frame 188\,$\mu$m, the extended dust continuum component is 
brighter than the compact dust component, while at the longer wavelength,  
rest-frame 419\,$\mu$m, the compact dust component is more luminous. 
This suggests that the compact dust continuum region has an intrinsic 
lower dust temperature than the extended dust component, and/or 
different submm optical depths. Such a picture 
is consistent with the merger scenario (but hard to be explained by 
dusty clumps in a single disk): the compact dust continuum component 
is tracing the large amount of cold dust peaking in the interacting 
region between R and B. This is similar to the cold dust maps seen in local mergers 
such as the Antennae (NGC\,4038/39) \citep[e.g.,][]{2000A&A...356L..83H, 
2000ApJ...542..120W} and VV\,114 \citep{2002A&A...391..417L}, where the cold 
dust emission at relatively long wavelengths is peaking in the interacting regions 
between two merging galaxies, suggesting a concentration of cold dust 
in a pre-coalescence phase of the system. While for the extended dust 
component, the contribution from the warm dust emission from the nucleus of 
R and B, elevates the overall dust temperature compared with the 
interacting region (i.e., the compact dust component). Therefore, the 
extended dust component has a warmer dust temperature compared with 
the compact dust component. The line emission of \co65\ and \htot211202\ 
is also predominantly coming from the warm dense regions (i.e., 
concentrated at R and B while remaining weak at the interacting region). 
However, the current spatial resolution limits our ability to resolve 
any further detailed variations in the properties of the dust emission. 
Higher spatial resolution and longer wavelength observations are needed 
to understand the distribution of the dust temperature and mass in  
R, B, and the interacting regions.

For the source-plane reconstruction of the \htop\ line emission, with 
only two velocity bins, it is difficult to infer the line's detailed 
spatial structure accurately. Nevertheless, we find no evidence that 
the spatial distribution of the \htop\ line emission is different from 
those of the \co65\ or \htot211202\ lines. This is consistent with the 
fact that the formation of \htop\ lines is associated with the strong 
cosmic rays from intensely star-forming regions, which are also traced 
by the \co65\ and \htot211202\ lines \citep{2016A&A...595A..80Y}. As in 
SMGs, which have very high-density star formation, we would expect to 
see the impact of cosmic rays on the ionization of dense molecular 
gas tracers as cosmic rays, rather uniquely, deposit their energy 
in deeply embedded dense gas \citep[e.g.,][]{2010ApJ...720..226P, 
2011MNRAS.414.1705P}.

\subsection{Kinematic structure}
\label{section:source_plane:kinematics}

\begin{figure*}[htbp]
\centering
\includegraphics[scale=0.694]{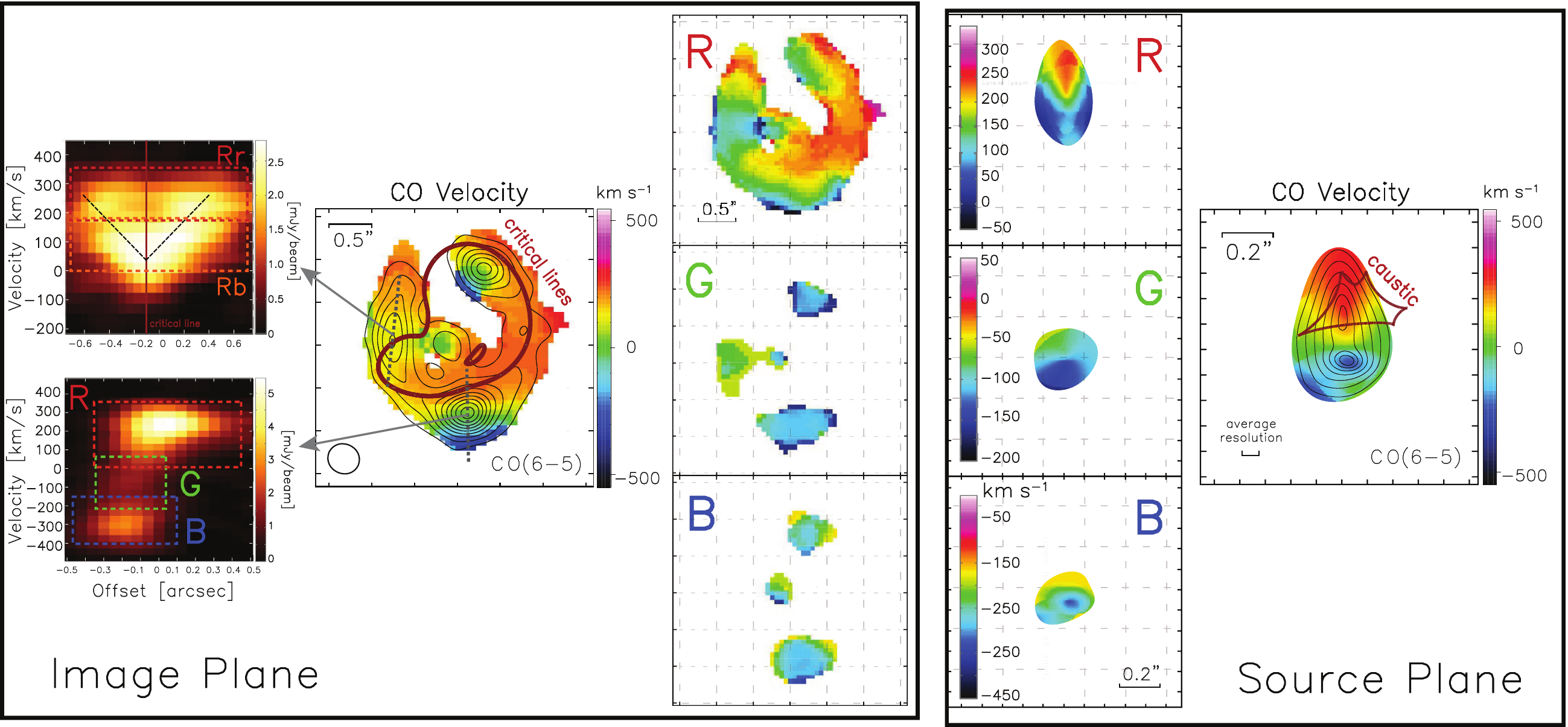}
\caption{
{\em Left panels, in the image plane}: 
PV plots in the image plane \co65\ data cube along the direction 
marked by the gray dashed lines are shown in the first column. The dark red line 
shows the position of the critical line. The dashed squares show the major 
velocity components ``R'' (Rb+Rr), ``G'', and ``B''. The second column 
shows the 0$^{\rm th}$ \& 1$^{\rm st}$ moment map of \co65\ similar to 
Fig.\ref{h2o:g09-mom-image}. The critical line is marked by dark red 
line. The third column shows the slice of 1$^{\rm st}$ moment CO maps 
according to the R, G, and B velocity ranges marked in the lower 
panel of the PV plot in the first column.
{\em Right panels, in the source plane}:
The first column shows the slice of 1$^{\rm st}$ CO moment maps for 
R, G, and B. The second column shows a similar 
image of 0$^{\rm th}$ \& 1$^{\rm st}$ moment maps of \co65\ 
in the image plane with the caustic line (dark red line) overlaid.
By comparing the velocity structures of R, G, and B both in the 
image plane and the source plane, and also the PV plot, it is clear that 
the source has three major velocity structures: the approaching gas B, 
the bridging gas G and the receding gas component R that has a 
velocity-resolved structure showing a velocity gradient along 
the north-south direction.
}
\label{fig:CO-PV}
\end{figure*}

Fig.\,\ref{fig:CO-PV} presents the source-plane reconstructed  
1$^{\rm st}$ moment map of the line emission and the same map 
in the image plane. The lens-model-reconstructed source-plane moment map  
shows the detailed intrinsic kinematic structure of G09v1.97. 
One should note that, as mentioned in Section\,\ref{section:lensmodel2}, 
because the final source-plane cubes are reconstructed by extracting the 
best fitted parametric lens models from the posteriors of the MCMC 
realizations per velocity channel, the noise of the observed images will 
not be transferred into the source-plane cubes. Thus, the uncertainties 
could be underestimated. For the dust continuum and line 
emission cubes, the noise in the ALMA images is generally low. 
Because of this, the uncertainties of the source-plane cubes are likely 
to be predominately due to the  
goodness of the lens model, namely the posterior distribution of the 
parameters, which reflects how well the model fits the data. 
This approach is a good approximation considering the limited spatial 
resolution yet high sensitivity levels of the dust continuum, \co65\  
and \htot211202\ line images. We will omit a discussion of 
the image plane map of \htot211202\ since the velocity structure 
is very similar to that of the \co65\ and we used fixed parameters 
when performing the lens modeling (Section\,\ref{section:lensmodel2}). 
With these caveats in mind, we find an overall velocity gradient of the gas 
along the north-south direction, with the northern gas component dominating
the velocity channels reward of the systemic velocity, 
while the southern component dominates the blue velocity channels,
This is in agreement with the R and B image-components discussed in 
Section\,\ref{section:source_plane:image}.

Although the 1$^{\rm st}$ moment map in the source plane has a 
rotation-like velocity shear, it remains incompatible with a single 
rotating-disk model (Section\,\ref{section:source_plane:image}).
This demonstrates the fact that a 2D velocity map with limited spatial 
resolution is generally insufficient to distinguish between a system that is a 
merger or a clumpy rotating disk.

To further investigate the velocity structure of G09v1.97, we extract 
PV plots sliced from two positions in the west and south of the 
image-plane map (Fig.\,\ref{fig:CO-PV}). The PV maps sliced from the 
west has a velocity range from $-$50 to 350\,\kms, which is mostly 
traces the northern disk R (which corresponds to the Rb and 
Rr components in the observed spectrum). It shows a typical mirrored folded 
structure along the two sides of the critical line (similar to e.g.,  
the Eyelash, \citealt{2011ApJ...742...11S}). The continuous velocity 
gradient along the positional axis of the PV diagram shows the typical 
kinematic signature of a rotating disk, suggesting that the 
northern galaxy is a kinematically resolved rotating disk. However, 
given the limited spatial resolution, we are unable to rule out the 
possibility that the R component can also be a small-scale 
merger \citep[see the model of e.g.,][]{2015Natur.525..496N}. 
We nevertheless assume R is a rotating disk in this work.
The other PV plot sliced from 
the southern component of the image shows the full velocity range from 
$-$400 to 350\,\kms, and which consists of three major components. The R part 
from 0\,\kms\ to 350\,\kms\ is tracing the same rotating disk structure 
as in the first PV plot (western slice), but is less magnified and therefore 
shows only a marginally resolved kinematic structure. The B part in the PV plot 
from $-$400 to $-$150\,\kms, which is associated with the southern 
galaxy, is coming from the B component in the 0$^{\rm th}$ 
moment map of the line emission. It is clear that the R and B  
components in the PV plot show spatially overlapping regions along 
the line of sight, suggesting that system cannot be a stable 
rotating disk. This again rules out the clumpy rotating disk
interpretation. In the overlap region, there is another weak gas 
component from $-$150 to 0\,\kms, probably tracing the bridging gas 
in between the two merging galaxies.

To test the velocity structure 
of these three components found in the southern PV plot, and to 
check the consistency between the source- and image-plane results, 
we have divided both the datacube in the source and image 
plane into three parts: R $-$400 to $-$150\,\kms, G $-$150 to 
0\,\kms, and B 0 to 350\,\kms. Fig.\,\ref{fig:CO-PV} shows the 
1$^{\rm st}$ moment maps for R, G, and B in the image and 
source plane. The component R shows a prominent velocity gradient 
consistent with a rotating disk, while G and B have  
no clear velocity structure. The moment maps from the image and source
plane match very well, suggesting that our lens model is robust.
The velocity structure further suggests that the system consists 
of two merging galaxies -- a northern
rotating disk with marginally resolved kinematics and a 
southern compact galaxy without any clear velocity 
structure -- during their pre-coalescence phase. These two 
galaxies are connected by a bridge of gas with very weak \co65\ 
emission and dominated by cold dust emission 
(see a sketch in Fig.\,\ref{fig:source-plane-spec}).

\subsection{The intrinsic properties of G09v1.97: ISM and star formation in the source plane}
\label{section:source_plane:sf}

\begin{figure*}[htbp]
\centering
\includegraphics[scale=1.02]{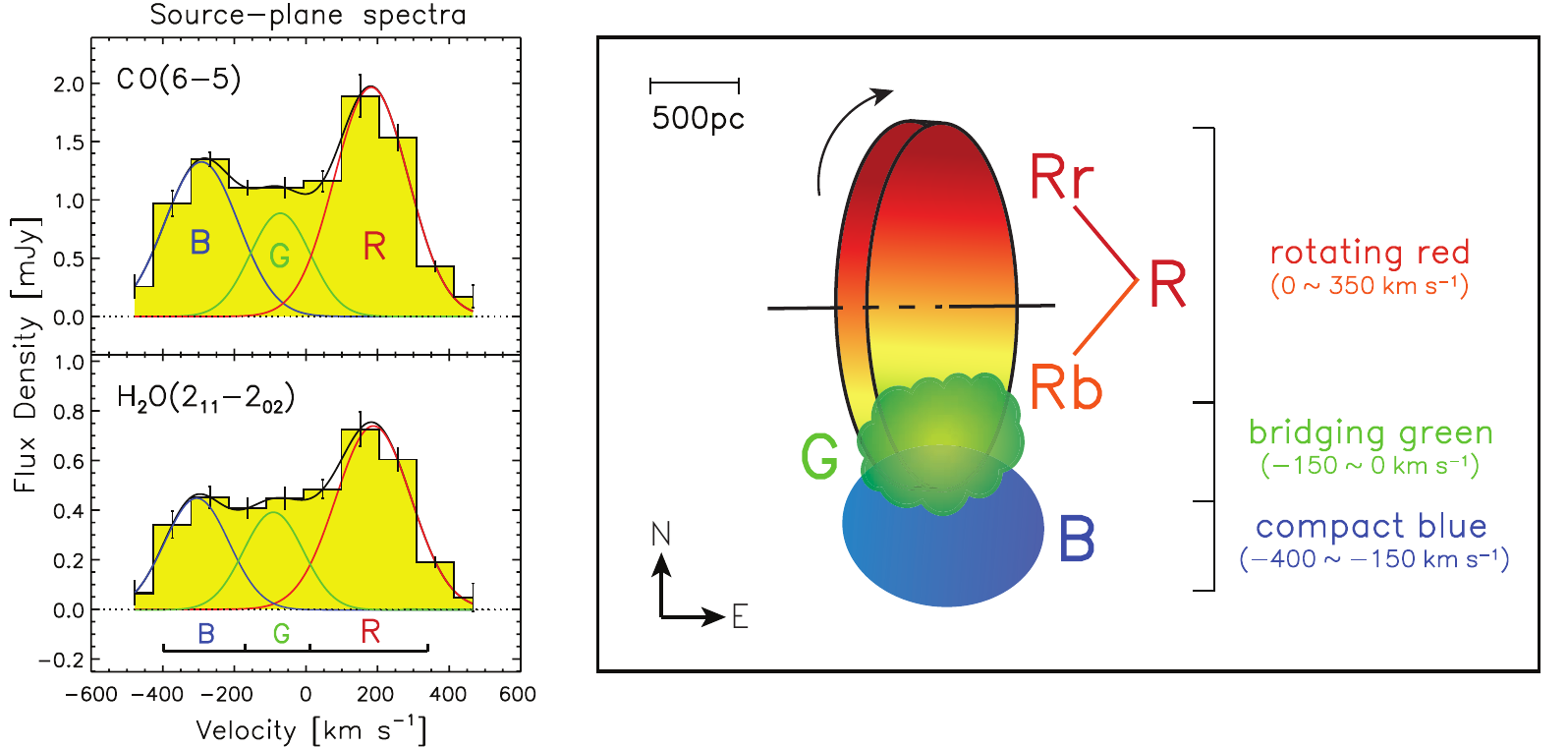}
\vspace{-0.2cm}
\caption{
{\em Left panel:} Spatially integrated spectra extracted from 
the source-plane line data-cubes constructed from the lens model. 
The error bars are dominated by the uncertainties in the data. 
The black curve shows the results of fitting 3 Gaussian line profiles: 
the blue line represents the B component, the green line 
represents the gas bridge G component, and the red line 
represents the receding gas R component (Rb+Rr). 
The line profiles of \co65\ and \htot211202\ are similar, 
having $\sim$\,1.7 times stronger (peak flux) red-shifted 
emission compared to the blue-shifted emission. This indicates 
that there is a similar spatial distribution between the \co65\ 
and \htot211202\ line emission. The asymmetry of the profile suggests 
an intrinsically complex velocity structure. 
{\em Right panel:} A sketch of the gas-rich major merger, G09v1.97. 
The source is composite of three major components, a rotating disk
to the north that corresponds to the R component of the spectra 
shown in the left panel or the double-Gaussian component Rb+Rr  
seen in the image-plane spectrum, while the compact blue galaxy in 
the south corresponds to B part of the line profile. There is 
also a dust-rich gas bridge between the two over the component which 
has weak \co65\ and \htot211202\ emission.
}
\label{fig:source-plane-spec}
\end{figure*}

We extract the spatially integrated spectrum of the \co65\ and \htot211202\ 
line from the reconstructed source-plane cubes, and further decompose the 
spectra into 3 Gaussian profiles (Fig.\,\ref{fig:source-plane-spec}).  
In the fitting process, we set all the parameters free, except for 
the linewidth of the G component of the \htot211202\ line (by using the 
value from the CO line) to achieve robust constraints on the parameters 
(Table\,\ref{table:src-spec}). Both the \co65\ and 
\htot211202\ line profiles are well-described by 3 Gaussians: 
a blue-shifted Gaussian profile located mainly between $-$400 and $-$150\,\kms\ 
(in terms of \FWHM), which is associated with the southern blue compact 
galaxy B; a red-shifted Gaussian component from 0 to 350\,\kms, 
originating in the northern rotating disk; and a weak Gaussian component 
from $-$150 to 0\,\kms\, dominated by the gas bridge connecting the two 
merging galaxies. It is noticeable that due to the differential lensing, 
the spectrum in the source plane shows a somewhat different line profile. 
The R component is heavily magnified by a factor of 15--20, and has been 
decomposed into two Gaussian components, Rb and Rr, which are the 
blue-shifted side and red-shifted side of the northern rotating disk. 
For the G component, due to its weak flux and lower magnification ($<$10), 
it is difficult to disentangle it from components Rb and B seen in 
the image-plane spectrum.

Comparing the linewidths of the red-shifted and blue-shifted Gaussian 
components of the \co65\ line and the \htot211202\ line, we find that 
the blue-shifted B has similar linewidth as the red-shifted R, 
despite the fact that the latter is resolved into a rotating disk 
while the former shows no clear velocity structure. This indicates 
B has a larger intrinsic velocity dispersion than R. The integrated 
flux ratio of R and B for the two lines are similar, 
$I_\mathrm{R}/I_\mathrm{B}$\,$\sim$\,1.5--2.1, with a peak flux ratio 
of about 1.6 for both. The bridging gas component contributes 20\% to 
the total flux for both the \co65\ and \htot211202\ lines. The flux 
ratios of \co65/\htot211202\ for R and B are within the typical 
flux ratios of \co65/\htot211202\ found in {\it H}-ATLAS SMGs 
\citep{2016A&A...595A..80Y, 2017A&A...608A.144Y}, and is slightly 
higher than the value $1.4\pm0.1$ found in Arp\,220 \citep{2011ApJ...743...94R}. 
However, this flux ratio of \co65/\htot211202\ is a slightly lower 
than the value of $2.5\pm0.7$ found in the similar IR-luminous $z$\,=\,5.2 
SMG HLSJ0918 \citep{2014ApJ...783...59R} and $3.4\pm1.0$ in the $z$\,=\,5.7 
SMG ADFS-27 \citep{2017ApJ...850....1R}. Since the excitation of the 
\htot211202\ line is dominated by far-IR pumping, it is tracing the 
warm dust emission \citep[with temperatures around 45\,K, see e.g.,][]{2014A&A...567A..91G}, 
while the \co65\ line is tracing the warm dense gas mass. The line ratio 
of CO to \hto\ may therefore offer an estimate of the gas-to-dust 
mass ratio in the warm dense star-forming molecular gas. If the flux 
ratio of \co65/\htot211202\ is indeed correlated with the gas-to-dust 
mass ratio, a similar ratio of \co65/\htot211202\ in R and B  
could indicate that the dust-to-gas mass ratios of the two merging 
galaxies are likely to be similar. While higher flux ratios of \co65/\htot211202\ 
in the $z$\,$>$\,5 sources, such as the previously discussed HLSJ0918 
and ADFS-27, suggests higher gas-to-dust mass ratios compared with 
the $z$\,=\,2--4 {\it H}-ATLAS SMGs and local ULIRGs. Although crude, 
this result is consistent with the fact that the higher redshift 
sources tend to be gas-rich and less enriched in metals and dust 
\citep[e.g.,][]{2015ApJ...800...20G}. More work with a larger 
sample is needed to test the reliability of such an assumption.

\begin{figure}[htbp]
\includegraphics[scale=0.7]{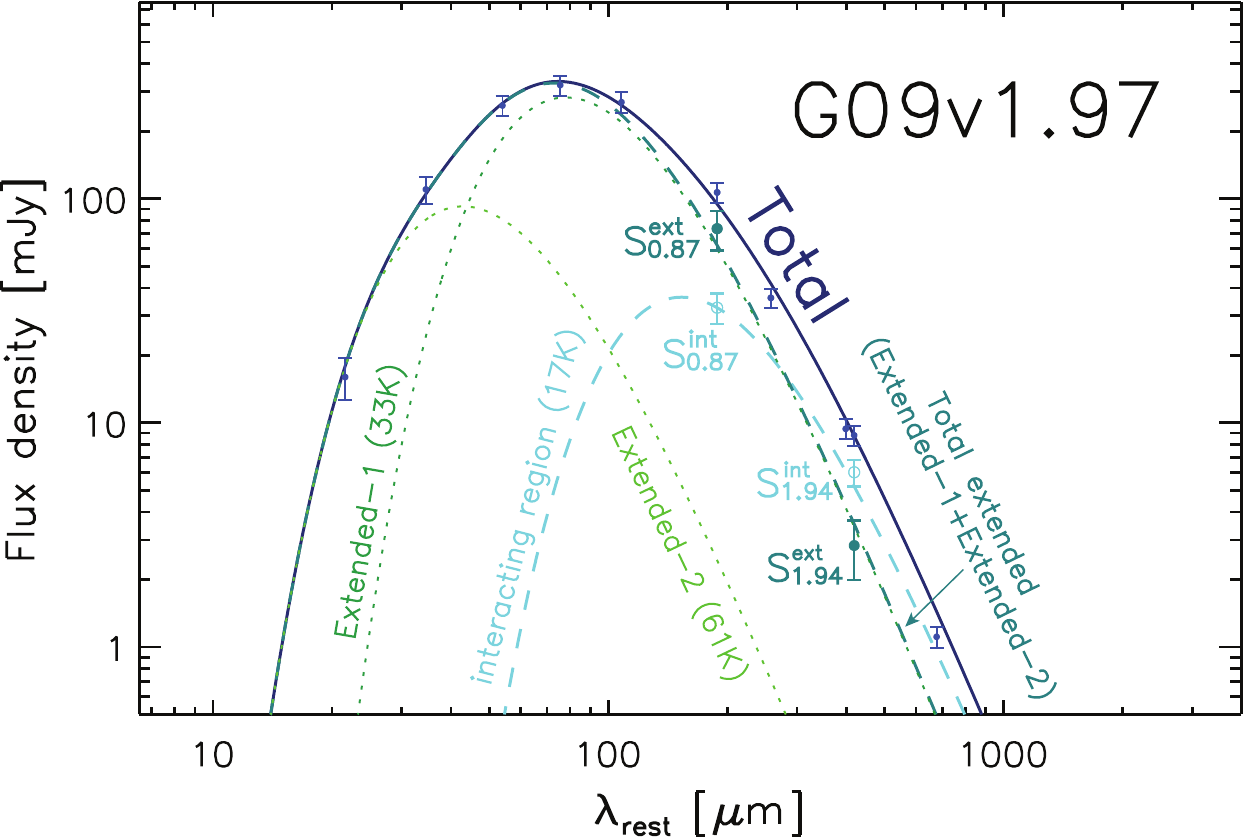}
\vspace{-0.2cm}
\caption{ 
Dust SED of G09v1.97. The fluxes are taken from {\it Herschel} 
and NOEMA measurements \citep{2016A&A...595A..80Y}, and also from 
this work for the 0.89 and 1.94\,mm continuum fluxes (blue points). 
The fit also takes into account the dust continuum fluxes of the 
compact (the interacting region) and extended (the merging galaxies) 
dust component from our lens model, indicated by the open and filled 
circle (green and light-blue for different temperature components), 
respectively. The SED fit has also taken the differential magnification 
for the compact and extended components into account. The solid blue 
line shows the best fit, while other dashed/dotted lines are 
decompositions of different temperature components. See text for details.  
}
\label{fig:dust_sed}
\end{figure}

The limited spatial resolution means that we are unable to resolve 
the detailed structure of the dust emission and disentangle 
the two galaxies and their interacting regions. Nevertheless, 
to first order, we can assume that the total infrared luminosity 
is proportional to the \co65\ luminosity based on the tight 
correlation found in infrared-bright galaxies \citep[e.g.,][]{2014ApJ...794..142G, 
2015ApJ...810L..14L,2017ApJS..230....1L}. Also, notice that 
the bridge has a lower dust temperature as indicated 
by its low ratio, $S_\mathrm{0.87}$/$S_\mathrm{1.94}$\,=\,5.4 
compared to R and B (for which we assume a similar ratio,  
$S_\mathrm{0.87}$/$S_\mathrm{1.94}$\,=\,26). The fluxes
of the two dust components at 0.87 and 1.94\,mm, and their
ratios are well constrained by our model, showing no sign of 
strong degeneracies between the parameters (Fig.\,\ref{fig:cornerplot}).
We can decompose the total \lir\ into three parts using a modified 
black-body model ($\beta$\,=\,2), for B, G, and R  
(Fig.\,\ref{fig:dust_sed}). We acknowledge that there could 
be large uncertainties because there is no spatially resolved 
photometry for B, G, and R. Therefore, we introduced 
additional uncertainties of 50\% to the resulting best-fit 
parameters.

\setlength{\tabcolsep}{0.68em}
\begin{table*}[htbp]
\small
\caption{Spatially integrated dust and line emission properties in the source plane.}
\begin{tabular}{llrrrccccc}
\toprule
 Line                        &  \llap{Co}mpon\rlap{ent} &$S_\mathrm{pk}$\;\;\;\; & $S_\mathrm{R}/S_\mathrm{B}$ & $I_\mathrm{line}$\,\;\;\;\; & $I_\mathrm{R}/I_\mathrm{B}$   & $I_\mathrm{G}/I_\mathrm{total}$ & \;\;$\Delta{V}_\mathrm{line}$  & $L_\mathrm{Line}/10^{7}$\;    & $ L'_\mathrm{Line}/10^{9}$ \\
                             &                          &     (mJy)\;\;\;        &                             & (Jy\,km\,s$^{-1}$)          &                               &                                 & (km\,s$^{-1}$)                 &  (\lsun)                      &       (\kkmspc)            \\
\midrule                                                                                                                                                                                                                                       
\multirow{3}{*}{\co65}       & B                        &\;\;\;  $1.3\pm0.6$     &\multirow{3}{*}{$1.5\pm0.7$} &       $0.3\pm0.1$           &\multirow{3}{*}{$1.5\pm0.5$}   &\multirow{3}{*}{$0.2\pm0.1$}     &   $243\pm65$                   &     $5.7\pm1.8$               &       $5.3\pm1.7$          \\
                             & G                        &\;\;\;  $0.9\pm0.8$     &                             &       $0.2\pm0.1$           &                               &                                 &   $197\pm90$                   &     $3.2\pm2.3$               &       $3.0\pm2.2$          \\
                             & R(Rb+Rr)                 &\;\;\;  $2.0\pm0.4$     &                             &       $0.5\pm0.1$           &                               &                                 &   $243\pm28$                   &     $8.5\pm1.3$               &       $8.0\pm1.3$          \\[.2cm]
\multirow{3}{*}{\htot211202} & B                        &\;\;\;  $0.4\pm0.3$     &\multirow{3}{*}{$1.7\pm1.2$} &       $0.1\pm0.1$           &\multirow{3}{*}{$2.1\pm1.2$}   &\multirow{3}{*}{$0.2\pm0.1$}     &   $212\pm67$                   &     $1.8\pm0.9$               &       $1.3\pm0.7$          \\
                             & G                        &\;\;\;  $0.4\pm0.2$     &                             &       $0.1\pm0.1$           &                               &                                 &   $197$\rlap{$^a$}             &     $1.4\pm0.7$               &       $1.1\pm0.5$          \\
                             & R(Rb+Rr)                 &\;\;\;  $0.8\pm0.3$     &                             &       $0.2\pm0.1$           &                               &                                 &   $261\pm40$                   &     $3.8\pm1.1$               &       $2.8\pm0.8$          \\
\midrule 
\multirow{3}{*}{\lir}        & B                        &                          \multicolumn{4}{c}{log(\lir/\lsun)\,=\,$12.6\pm0.3$}                                      &                               \multicolumn{3}{c}{\td\,=\,{$33\text{--}61$}\,K}                           &                            \\
                             & G                        &                          \multicolumn{4}{c}{log(\lir/\lsun)\,=\,$11.6\pm0.3$}                                      &                               \multicolumn{3}{c}{\td\,=\,{$17^{+9}_{-4}$}\,K}                        &                            \\
                             & R(Rb+Rr)                 &                          \multicolumn{4}{c}{log(\lir/\lsun)\,=\,$12.8\pm0.3$}                                      &                               \multicolumn{3}{c}{\td\,=\,{$33\text{--}61$}\,K}                           &                            \\
\bottomrule
\end{tabular}
\tablefoot{ The fitting errors are dominated by the uncertainties in lens 
modeling. The uncertainties of dust far-IR SED fitting is significant due to the 
limited constraints and the complexities of lens modeling. Therefore, we increase 
the uncertainties by 50\% for the total IR luminosity and dust temperature. For 
separating the \lir\ for different component, we assume a similar ratio 
for \lir\ as found in $L_\mathrm{CO(6\text{--}5)}$ for R and B, and the 
dust temperatures were derived based on a global dust SED fitting.
$^{(a)}$: The linewidth has been fixed for G using those 
determined from fitting the CO line over the same emission region.
}
\label{table:src-spec}
\end{table*}
\normalsize

The two merging galaxies are infrared luminous, with 
\lir$_\mathrm{,\,B}$\,=\,(4.0$\pm$2.0)$\times$10$^{12}$\,\lsun\ 
and \lir$_\mathrm{,\,R}$\,=\,(6.3$\pm$3.1)$\times$10$^{12}$\,\lsun, 
and can thus be considered as ULIRGs. The dust emission between the two galaxies has a
\lir\,=\,(4.0$\pm$2.0)$\times$10$^{11}$\,\lsun. The merging system 
has a total \lir\ exceeding 10$^{13}$\,\lsun\ and therefore belongs 
to the class of hyper-luminous infrared galaxies (HyLIRGs). The 
intrinsic dust temperature for the two galaxies is about 33--61\,K 
(composite of two temperature components of 33 and 61\,K), which 
is typical for high-redshift SMGs \citep[e.g.,][]{2005ApJ...622..772C, 
2012A&A...539A.155M, 2016ApJS..222....4S}. The region between the 
two ULIRGs has a much lower dust temperature of 17\,K. This 
temperature is consistent with values found for the coldest dust 
component in a sample of high-redshift SMGs \citep{2012A&A...539A.155M}.
The corresponding star formation rates for B and R are 
$\mathit{SFR}_\mathrm{B}$\,=\,(6$\pm$3)$\times$10$^{2}$\,\msun\,yr$^{-1}$ 
and $\mathit{SFR}_\mathrm{R}$\,=\,(9$\pm$4)$\times$10$^{2}$\,\msun\,yr$^{-1}$.
Taking the intrinsic half-light radii of 0.4\,kpc and 1.2\,kpc for the 
two galaxies, we derive the surface star formation intensities, 
$\Sigma_\mathit{SFR,\mathrm{B}}$\,=\,(5.5$\pm$2.8)$\times$10$^{2}$\,\msun\,yr$^{-1}$\,kpc$^{-2}$ 
and $\Sigma_\mathit{SFR,\mathrm{R}}$\,=\,(1.2$\pm$0.6)$\times$10$^{2}$\,kpc$^{-2}$ 
for the B and R galaxies, respectively. Comparing these surface 
intensities with theoretical limits derived for an optically thick disk, where 
the radiation pressure on dust grains provides the pressure to support a disk 
against its own self-gravity \citep{2005ApJ...630..167T}, the values of the 
star formation intensities are 2--7 times smaller and thus well below this 
``Eddington limit''. However, the two components of G09v1.97 are still among 
the strongest starbursts found in {\it H}-ATLAS sample of lensed SMGs in terms 
of their $\Sigma_\mathit{SFR}$ \citep{2013ApJ...779...25B} and also comparable 
with SPT-selected sources \citep{2016ApJ...826..112S}. Note that the 
$\Sigma_\mathit{SFR}$ will be further reduced by a factor of up to 5--7, if we 
allow for a top-heavy IMF \citep{2018Natur.558..260Z}.

\begin{figure}[!htbp]
\includegraphics[scale=0.64]{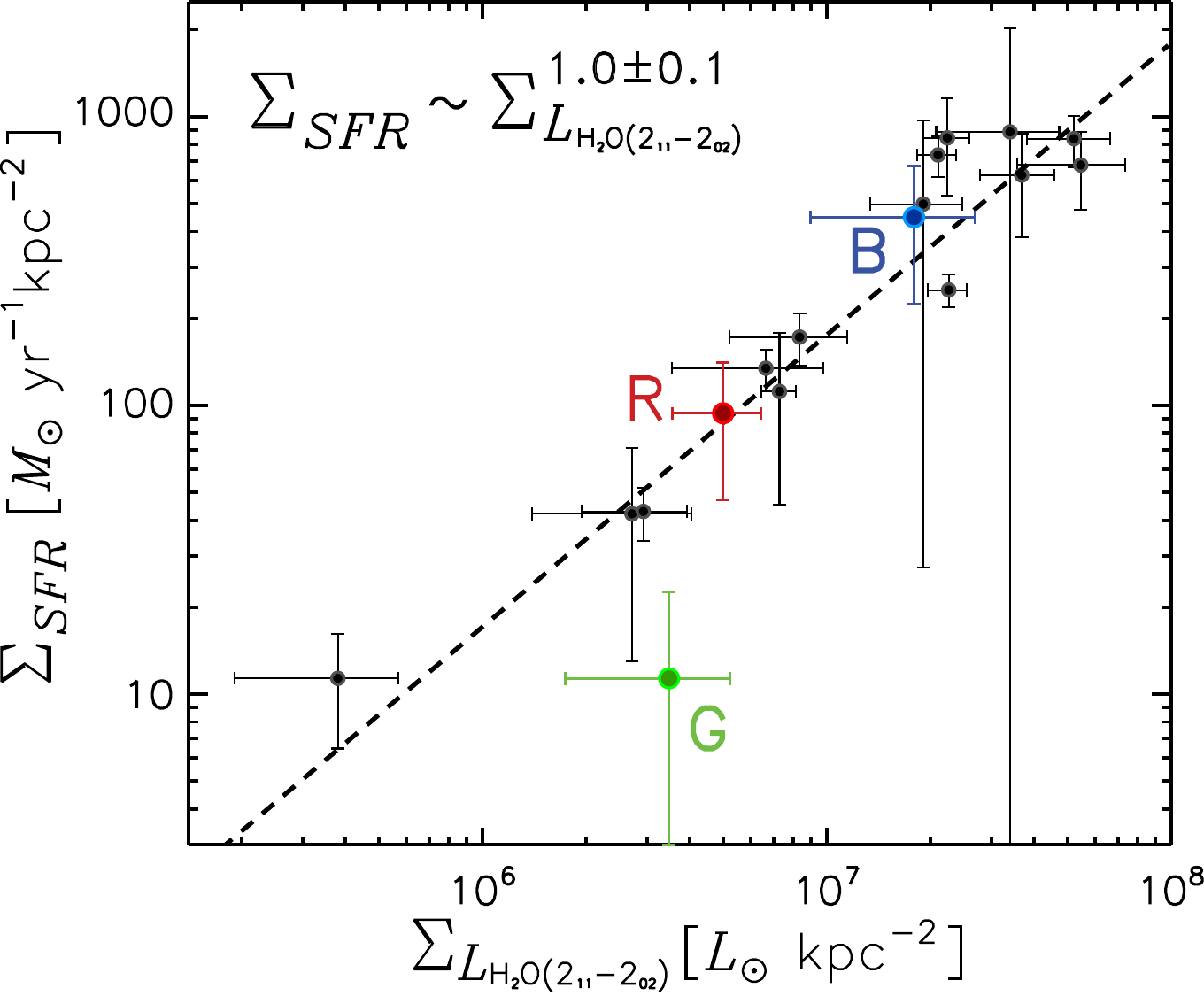}
\vspace{-0.2cm}
\caption{ 
Correlation between the star formation rate (SFR) surface 
density and \hto\ luminosity surface density in high-redshift SMGs 
\citep{2016A&A...595A..80Y} (shown in black), including HFLS3 
\citep{2013Natur.496..329R} and ADFS-27 \citep{2017ApJ...850....1R}. 
The red, blue and green data points represent the R, B, and G  
components of G09v1.97 from this work. The figure shows a tight 
linear correlation between $\Sigma_{\mathit SFR}$ and 
$\Sigma_{\mathit L_{\rm H_{2}O}}$ with a fitted slope of $1.0\pm0.2$, 
which is valid down to the sub-kpc scales. 
}
\label{fig:g09-l-l}
\end{figure}

Assuming similar gas excitation conditions, i.e., taking the same \co65\ 
to \co10\ ratio for R and B as in \cite{2017A&A...608A.144Y},
we can derive the total molecular gas mass (acknowledging the large 
uncertainties) for R and B of about 5$\times$10$^{10}$ and 
4$\times$10$^{10}$\,\msun, respectively, suggesting the two galaxies are 
gas-rich. The implied gas depletion time then is 60 and 63\,Myr 
for R and B, respectively, which are typical values found for 
high-redshift dusty starbursts \citep[e.g.,][]{2016ApJ...833...68A}. 
When comparing the luminosity ratio of \co65\ and infrared, we find 
values of \lco65/\lir\,=\,(1.0$\pm$0.5)$\times$10$^{-5}$ for the 
southern galaxy B and (2.1$\pm$1.1)$\times$10$^{-5}$ for the northern 
galaxy R, which are well within the range found by statistically 
studying other \hz\ SMGs \citep{2017A&A...608A.144Y}, and very close 
to the ratios of the local ULIRGs whose emission is dominated by 
star formation, not AGN \citep{2015ApJ...810L..14L}. For the \htot211202\ 
line, we derive a ratio \lhtot211202/\lir\ of (0.5$\pm$0.3)$\times$10$^{-5}$ 
and (0.6$\pm$0.3)$\times$10$^{-5}$ for B and R, respectively. 
These values are also in agreement with that of other SMGs and 
local star-forming dominated ULIRGs \citep{2016A&A...595A..80Y, 
2017ApJ...846....5L}, showing that the tight correlation still holds 
down to sub-kpc scales in such starbursts (Fig.\ref{fig:g09-l-l}), 
further suggests that the submm \hto\ lines are robust indicators 
of star formation rates in these galaxies. In Fig.\ref{fig:g09-l-l}, 
the G component has a small positional offset, showing an excess of \hto\ 
line emission. Nevertheless, given the fact that the estimate of the 
SFR is very uncertain in the interaction region, it is difficult to 
conclude that such a deviation is significant. However, shocks 
over the interaction region between two merging galaxies could also 
enhance low-$J$ \hto\ lines which may explain an excess of 
\hto\ luminosity \citep{2010MNRAS.406.1745F, 2013ApJ...777...66A}. 
Higher resolution observations are needed to further investigate 
such a hypothesis.

Our new ALMA data indicate that the ratio of the observed \ihtot211202\ 
to $I_\mathrm{H_2O^+}$ (the 742.1\,GHz line) is about 3 for the merging galaxies   
(using a total averaged magnification for \htop), which agrees with the 
\ihto/$I_\mathrm{H_2O^+}$ ratios reported in \cite{2016A&A...595A..80Y} for 
NCv1.143, G09v1.97, and G15v2.779 based on NOEMA compact array observations. 
With this ratio, the model discussed in \citet{2011A&A...525A.119M}, suggests 
a high cosmic-ray ionization rate of 10$^{-14}$--10$^{-13}$\,s$^{-1}$, with the 
upper limit being similar to the values derived in Arp\,220 and NGC\,4418 
\citep{2013A&A...550A..25G}. Such high cosmic-ray ionization rates drive the 
level of the ambient ionization of the high column density clouds to $\sim$\,10$^{-3}$, 
rather than the canonical 10$^{-4}$. These cosmic rays, which are likely generated 
in the most intense star-forming regions, dominate the gas phase ionization and 
oxygen ion-neutral route of the oxygen 
chemistry \citep[e.g.,][]{2011A&A...525A.119M, 2016A&A...593A..43V}.

In addition to the \htop\ lines, we also detected an emission line around 
745.3\,GHz, which we attribute to the H$_2^{18}$O(\t211202) line 
emission ($E_\mathrm{up}$\,$\sim$\,200\,K). The oxygen isotope, $^{18}$O, 
is predominantly made in the early stages of 
helium burning in massive stars \citep[e.g.,][]{1983A&A...120..113M}. H$_2^{18}$O 
has also been detected in emission in local ULIRGs both in a local {\it Herschel} 
archive survey \citep{2013ApJ...771L..24Y} and in the compact obscured nucleus 
of Zw\,049.057 \citep{2015A&A...580A..52F}. Several absorption lines of this 
molecule at shorter wavelengths have also been detected by {\it Herschel}/PACS 
in local ULIRGs \citep{2012A&A...541A...4G, 2018ApJ...857...66G}, indicating 
that the excitation of H$_2^{18}$O is similar to that of \hto. The integrated 
flux ratio of \htot211202\ to H$_2^{18}$O(\t211202) is $\sim$\,15, while 
the peak flux ratio is about 8. However, with only one transition line of 
H$_2^{18}$O detected, it is very difficult to derive the 
abundance ratio of $^{18}$O/$^{16}$O because of the very large optical depth 
of the \hto\ line. With better constraints on the H$_2^{18}$O abundance (and 
careful astrochemical modeling), we can use this ratio to gain further insight 
into the IMF in \hz\ SMGs because $^{18}$O is a secondary nuclide that 
relies on the prior existence of the primary elements such as $^{12}$C and $^{16}$O 
\cite[e.g.,][]{1968psen.book.....C, 2013MNRAS.436.2793D, 2017MNRAS.470..401R, 2018Natur.558..260Z}.

\section{Summary}
\label{section:conclusions}

In this paper, we have presented high (0\farcs2--0\farcs4) angular resolution  
ALMA observations of the lensed redshift $z$\,=\,3.632 SMG, G09v1.97, in 
the dust continuum at the rest-frame wavelengths of 188 and 419\,\mum\ 
and in the molecular line emission of \co65, \htot211202, and $J_{\rm up}$\,=\,2 \htop. 
The 2\,mm spectra provide accurate values for the apparent luminosity of the lines 
and detailed line profiles, as well as the first detection of 
a line of $\rm H_2^{18}O$ at high redshift. The images of the dust 
continuum and line emission show a nearly complete 1\farcs5 diameter 
Einstein ring with substructures and a weaker emission peak at the center. 
The central image is the result of the rare lens configuration of two 
compound foreground deflectors. With a magnification of $\mu$\,$\sim$\,5--22 
for the line across different velocity channels, 
and $\mu$\,$\sim$\,10--11 for the continuum, the average boosted 
angular resolutions can reach down to scales of $\sim$\,0.4--1.3\,kpc, 
resulting in one of the highest angular resolution images of water 
line emission in an extra-galactic source to date.

With careful lens modeling, we have reconstructed the source-plane images of 
G09v1.97 in the dust continuum and in the molecular emission lines of \co65,
\htot211202, and \htop. The model reproduces, in the image plane, most of 
the details of the observed images of the dust continuum and the 
three-dimensional line emission data cubes. The dust emission appears 
as a compact disk surrounded by an extended overlapping disk structure. 
The two dust components have slightly different dust temperatures and/or 
submm optical depths, suggesting the dust properties in its peak region  
and the extended regions are different. For the molecular line emission, 
we find very similar kinematic structure and spatial distribution between 
the \co65\ and \htot211202\ lines. The line emission shows a bimodal 
structure, with a northern disk-like gas component associated with 
red-shifted gas emission while the southern compact component is dominated 
by the blue-shifted gas. The cold dust emission peaks in between the 
two gas components. We argue that the two gas components in the south and 
north are two gas-rich galaxies, each having a total molecular gas 
exceeding 10$^{10}$\,\msun. They are in the pre-coalescence phase, with 
a large amount of cold dust presented in their interacting region, and 
also weak bridging gas emission. The northern galaxy is kinematically 
resolved as a rotating disk with a semi-major half-light radius $a_s$ 
of 1.2\,kpc, while the southern galaxy, kinematically unresolved by being 
more turbulent, is compact with $a_s$\,=\,0.4\,kpc traced by the \co65\ line. 
The ratios of the total IR luminosities \lir\ to the \co65\ and 
\htot211202\ line luminosities in the two merging galaxies are similar 
and consistent with the values found in other \hz\ SMGs, supporting our 
contention that there is a tight correlation between star formation and 
the \hto\ line luminosity. The \co65/\htot211202\ flux ratios 
are found to be $\approx$2 (with 20\% uncertainties), which is consistent 
with the $z$\,=\,2--4 SMGs, while this flux ratio is found to be higher in 
$z$\,$>$\,5 SMGs. The detection of strong \htop\ emission and its relative 
strength to the \hto\ line suggests that the cosmic ray plays an 
important role in the regulation of the physical and chemical processes in 
the ISM of high-redshift SMGs. The star formation rate surface densities 
of the two components are in the range of $\sim$\,120--550\,\msun\,yr$^{-1}$\,kpc$^{-2}$,
being $\sim$\,2--7 times below the value set by the Eddington limit, yet 
they are among the strongest starbursts at high redshift.

In addition to the \co65\ and \htot211202\ line, we also present the first
5-$\sigma$ detection of H$_2^{18}$O(\t211202), which is about 8 
times weaker than the \htot211202\ line. This molecule may offer 
important information on the chemical processes of the ISM and 
the initial mass function in high-redshift galaxies.

Future higher angular resolution observations will be needed to explore
and trace in greater detail the distribution and kinematics of the gas
content in G09v1.97. With its very high luminosity, the gas-rich 
merger, G09v1.97, remains an exceptional target for further studies
at high spatial resolution. For instance, with ALMA, a resolution
of 0\farcs05 would help to further resolve the source at scales 
below 100\,pc for G09v1.97, which is about the characteristic 
sizes of associations of molecular clouds. Such high spatial resolution 
observations will allow us to probe the properties of distinct molecular 
clumps in this star-bursting galaxy in the early universe.

\begin{acknowledgement}
We thank the anonymous referee for very helpful comments and suggestions.
C.Y. was supported by an ESO Fellowship. C.Y. thanks Johan Richard and 
Martin Bureau for insightful discussions, and also thanks John Carpenter 
and Edward Fomalont for discussion on ALMA data reduction. 
This paper makes use of the following ALMA data: ADS/JAO.ALMA\#2015.1.01320.S 
and \#2013.1.00358.S. ALMA is a partnership of ESO (representing its 
member states), NSF (USA) and NINS (Japan), together with NRC (Canada), 
MOST and ASIAA (Taiwan), and KASI (Republic of Korea), in cooperation 
with the Republic of Chile. The Joint ALMA Observatory is operated by 
ESO, AUI/NRAO, and NAOJ. 
C.Y. and Y.G. acknowledge support by National Key R\&D Program of China 
(2017YFA0402700) and the CAS Key Research Program of Frontier Sciences. 
R.G., C.Y., and A.O. acknowledge the Programme National Cosmology and 
Galaxies for financial support in the early stages of this project. 
C.Y., A.O., and Y.G. acknowledge support from the NSFC grants 11311130491 
and 11420101002. C.Y., A.O., A.B., and Y.G. acknowledge support from 
the Sino-French LIA-Origins joint exchange program. 
I.R.S. and A.M.S. acknowledge support from STFC (ST/P000541/1).
S.D. is supported by the UK STFC Rutherford Fellowship scheme.
E.I.\ acknowledges partial support from FONDECYT through grant 
N$^\circ$\,1171710. 
E.G.-A. is a Research Associate at the Harvard-Smithsonian 
Center for Astrophysics, and thanks the Spanish Ministerio 
de Econom\'{\i}a y Competitividad for support under 
projects FIS2012-39162-C06-01 and ESP2015-65597-C4-1-R, 
and NASA grant ADAP NNX15AE56G. RJI acknowledges support 
from ERC in the form of the Advanced Investigator Programme, 
321302, COSMICISM. 
H.F. acknowledges support from NSF grant AST-1614326. 
D.R. acknowledges support from the National Science 
Foundation under grant number AST-1614213.
M.N. acknowledges financial support from the European 
Union's Horizon 2020 research and innovation programme 
under the Marie Sk\l{}odowska-Curie grant agreement No 707601.
M.J.M. acknowledges the support of the National Science Centre, 
Poland, through the POLONEZ grant 2015/19/P/ST9/04010;
this project has received funding from the European 
Union's Horizon 2020 research and innovation programme under 
the Marie Sk{\l}odowska-Curie grant agreement No. 665778.
I.P.-F. acknowledges support from the Spanish grants 
ESP2015-65597-C4-4-R and ESP2017-86852-C4-2-R.
US participants in \hatlas\ acknowledge support 
from NASA through a contract from JPL. Italian participants 
in \hatlas\ acknowledge a financial contribution from the 
agreement ASI-INAF I/009/10/0. SPIRE has been developed by 
a consortium of institutes led by Cardiff Univ. (UK) and 
including: Univ. Lethbridge (Canada); NAOC (China); CEA, 
LAM (France); IFSI, Univ. Padua (Italy); IAC (Spain); Stockholm 
Observatory (Sweden); Imperial College London, RAL, UCL-MSSL, 
UKATC, Univ. Sussex (UK); and Caltech, JPL, NHSC, Univ. Colorado (USA). 
This development has been supported by national funding agencies: 
CSA (Canada); NAOC (China); CEA, CNES, CNRS (France); ASI (Italy); 
MCINN (Spain); SNSB (Sweden); STFC, UKSA (UK); and NASA (USA). 
\end{acknowledgement}

\vspace{-0.5cm}

\small
\bibliographystyle{aa_url}
\bibliography{ref}

\normalsize
\onecolumn

\appendix

\section{Details on the lens model}
\label{appendix:models}
\begin{figure}[htbp]
\centering
\includegraphics[width=0.96\textwidth]{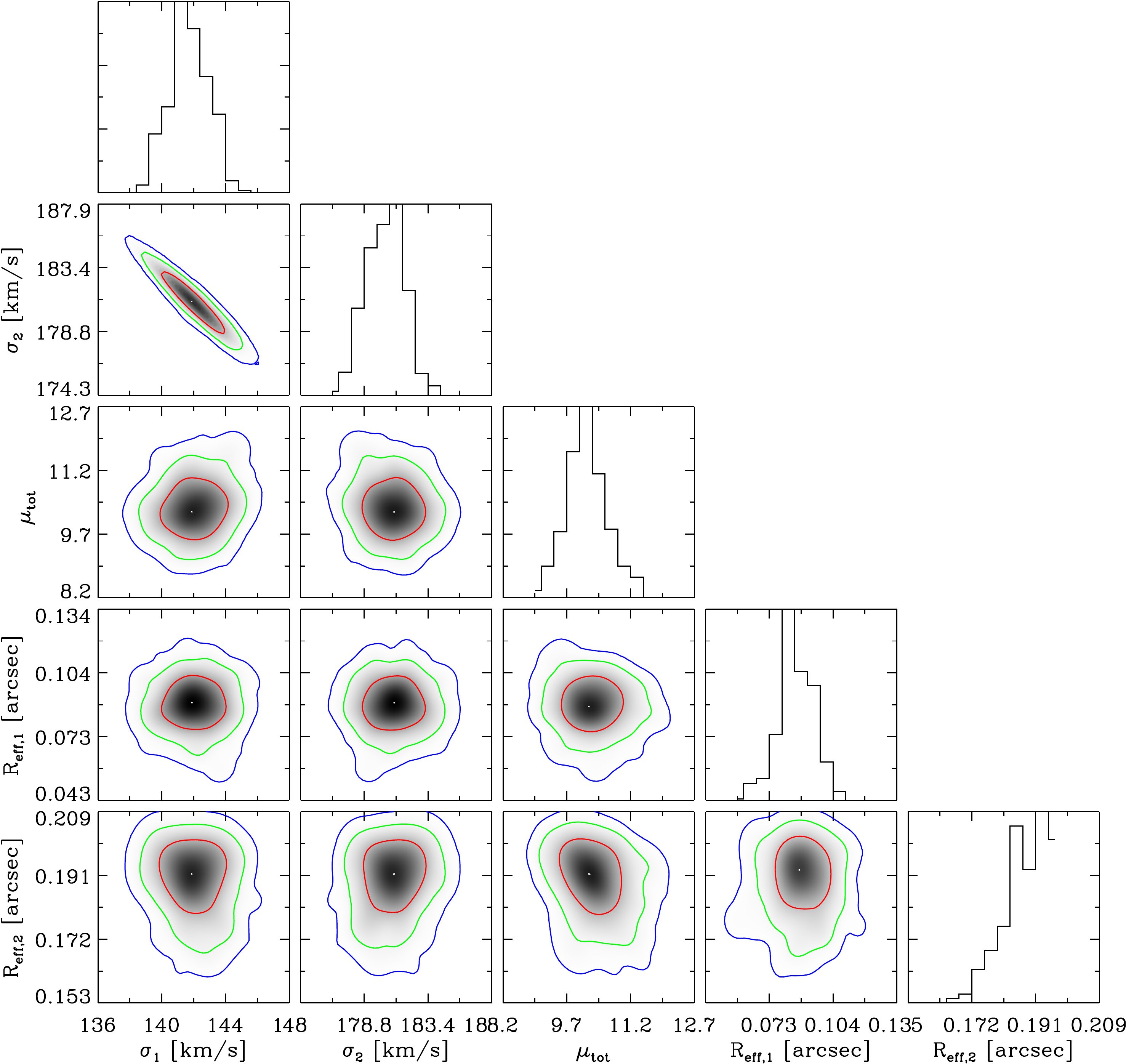}
\caption{Constraints on some relevant model parameters characterizing the two exponential profiles used to describe the
continuum emission in terms of half-light radius ($R_{\rm eff,1}$ and $R_{\rm eff,2}$),
total magnification factor ($\mu_{\rm tot}$) and stellar velocity dispersion (strength of the SIS mass distribution) for the two deflectors. For each component, magnification and size are somewhat degenerate 
with the flux density. Contours are 1-, 2-$\sigma$, and 3-$\sigma$ CL regions. Vertical lines in marginal PDF 
panels along the diagonal represent the mean and the $\pm1 \sigma$ interval around the mean.}
 \label{fig:cornerplot}
 \end{figure}

In addition to the model parameters listed in Table\,\ref{tab:lensmodel-G09}, we show in 
Fig.\,\ref{fig:cornerplot}, the constraints (pair-wise covariance and marginal 
probability density functions, i.e., PDFs) we derived for some relevant parameters 
defining the continuum flux, size, and magnification of the two exponential 
profiles.

Note that the uncertainties of $\mu$ (and on the intrinsic source flux and size) do 
not account for the effect of mass-sheet degeneracy and do not fully explore the 
degeneracy of these inferred parameters with the density profile slope of the mass 
distribution. This fundamental limitation of lens modeling at recovering intrinsic 
source-plane quantities are generally overlooked. However, the assumption of  
isothermal mass distribution, which is able to break somewhat the degeneracy, 
is largely confirmed by observations \citep{2009ApJ...703L..51K,2013ApJ...777...98S} 
and cosmological simulations \citep{2013MNRAS.433.3297D,2017MNRAS.472.2153P, 
2017MNRAS.469.1824X,2017MNRAS.464.3742R}.

\begin{figure*}[htbp]
	 \begin{center}
	    \includegraphics[width=0.985\textwidth]{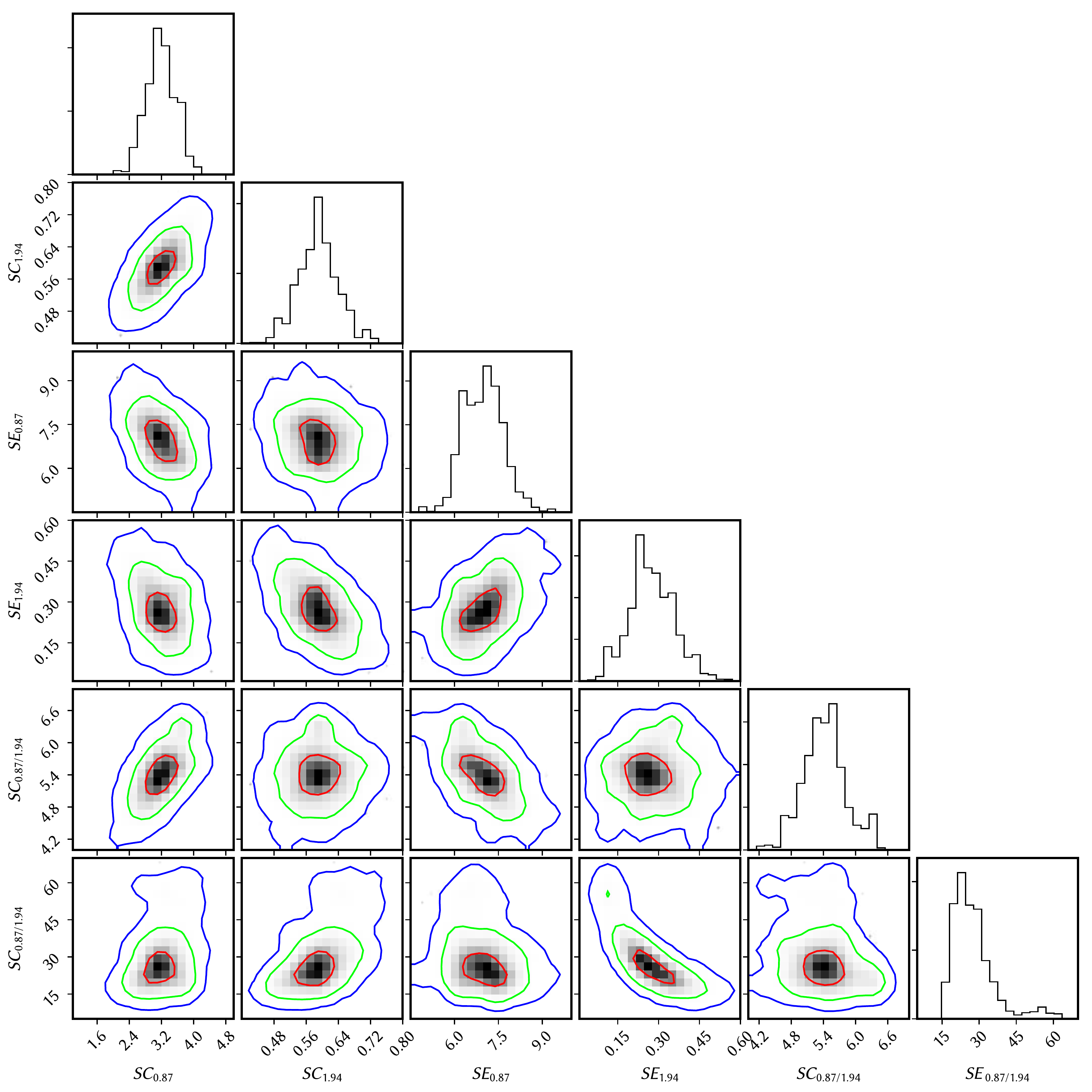}
     \end{center}
     \small{\textbf{Fig.~\ref{fig:cornerplot}.} Continued, for the flux densities of the compact and 
     extended dust component at 1.94 ($S\!C_\mathrm{1.97}$, $S\!E_\mathrm{1.97}$) and 0.87\,mm 
     ($S\!C_\mathrm{0.87}$, $S\!E_\mathrm{0.87}$). The unit is mJy. And the flux ratios of the  
     two bands for the compact ($S\!C_\mathrm{0.87/1.97}$) and extended ($S\!E_\mathrm{0.87/1.97}$) 
     dust component are also included.
     }
 \end{figure*}

 \begin{figure*}[htbp]
	 \begin{center}
	  \includegraphics[width=0.49\textwidth]{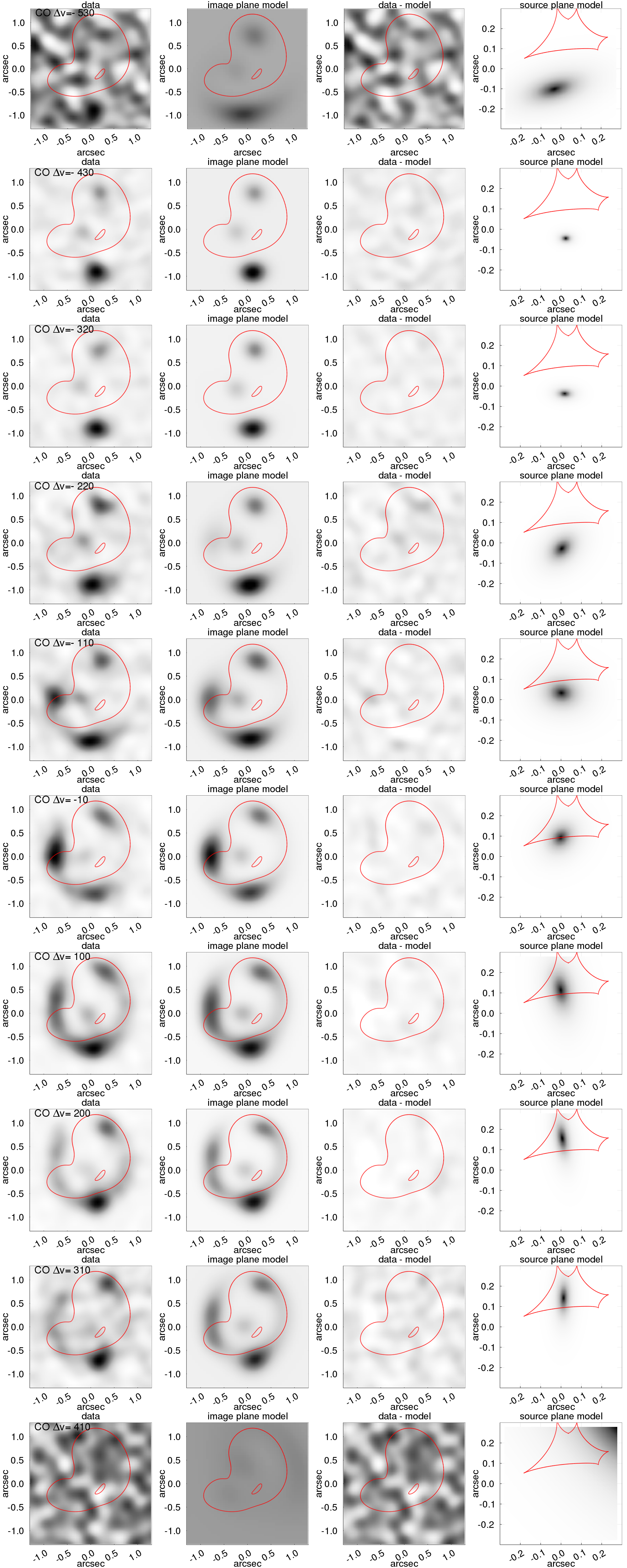}\ \ \ 
      \includegraphics[width=0.49\textwidth]{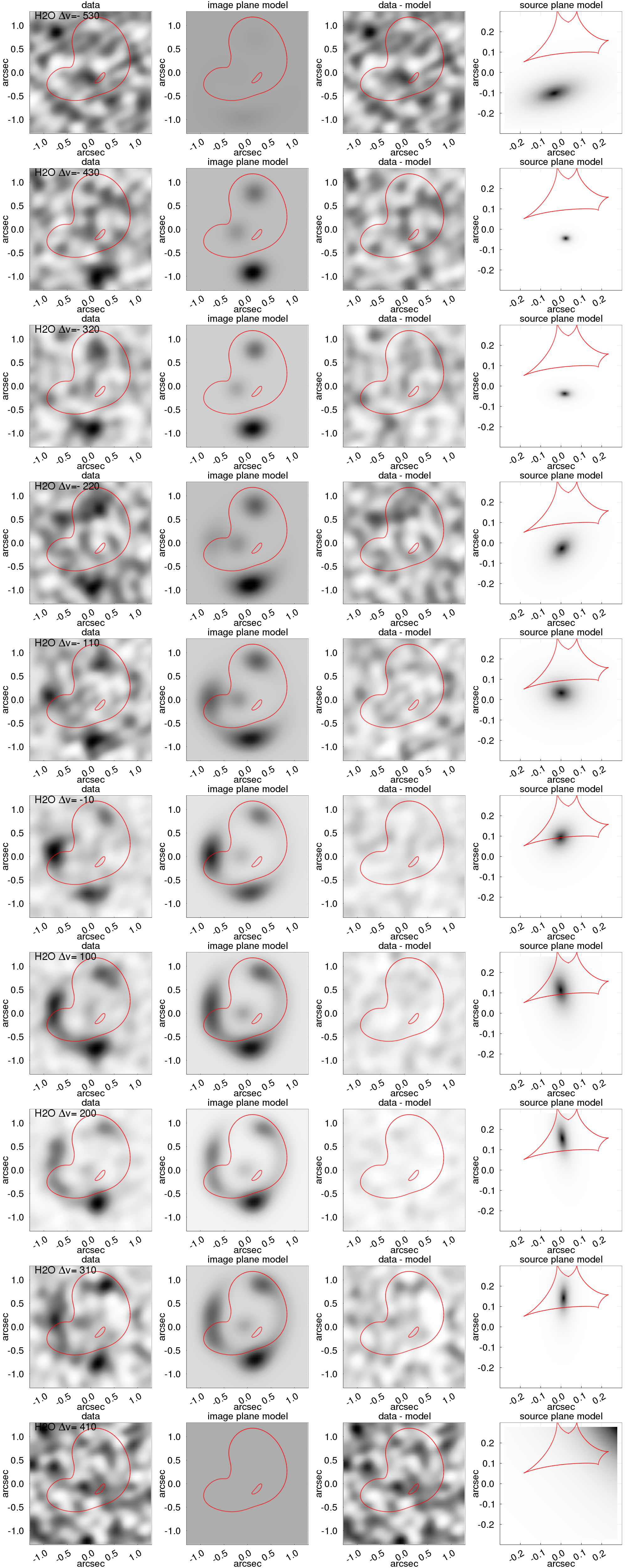} 
	\caption{Coarse-grained channel-by-channel comparison of the reconstruction of line emissions. The content of each sub-panel is like in Fig.~\ref{fig:resid-g97}. From left to right, it shows the observed data (image plane), model-predicted image plane reconstruction, difference of these images, and right, model-predicted source plane reconstruction, successively for the \co65\ and \htot211202\ lines. Sub-panels in columns (1,2,3,5,6,7) exhibit critical lines whereas the ones in columns (4,8) show caustic lines. From top to bottom, line emission channels are ordered with decreasing radial velocity, starting from $-$530\,\kms\ (the beginning velocity of each bin), in steps of 105\,\kms. Throughout those channel maps, no residual above the $\pm 3\sigma$ limit are found.}
 		\label{fig:allvelchannels}
     \end{center}
 \end{figure*}

 \begin{figure*}[htbp]
	 \begin{center}
	    \includegraphics[width=0.53\textwidth]{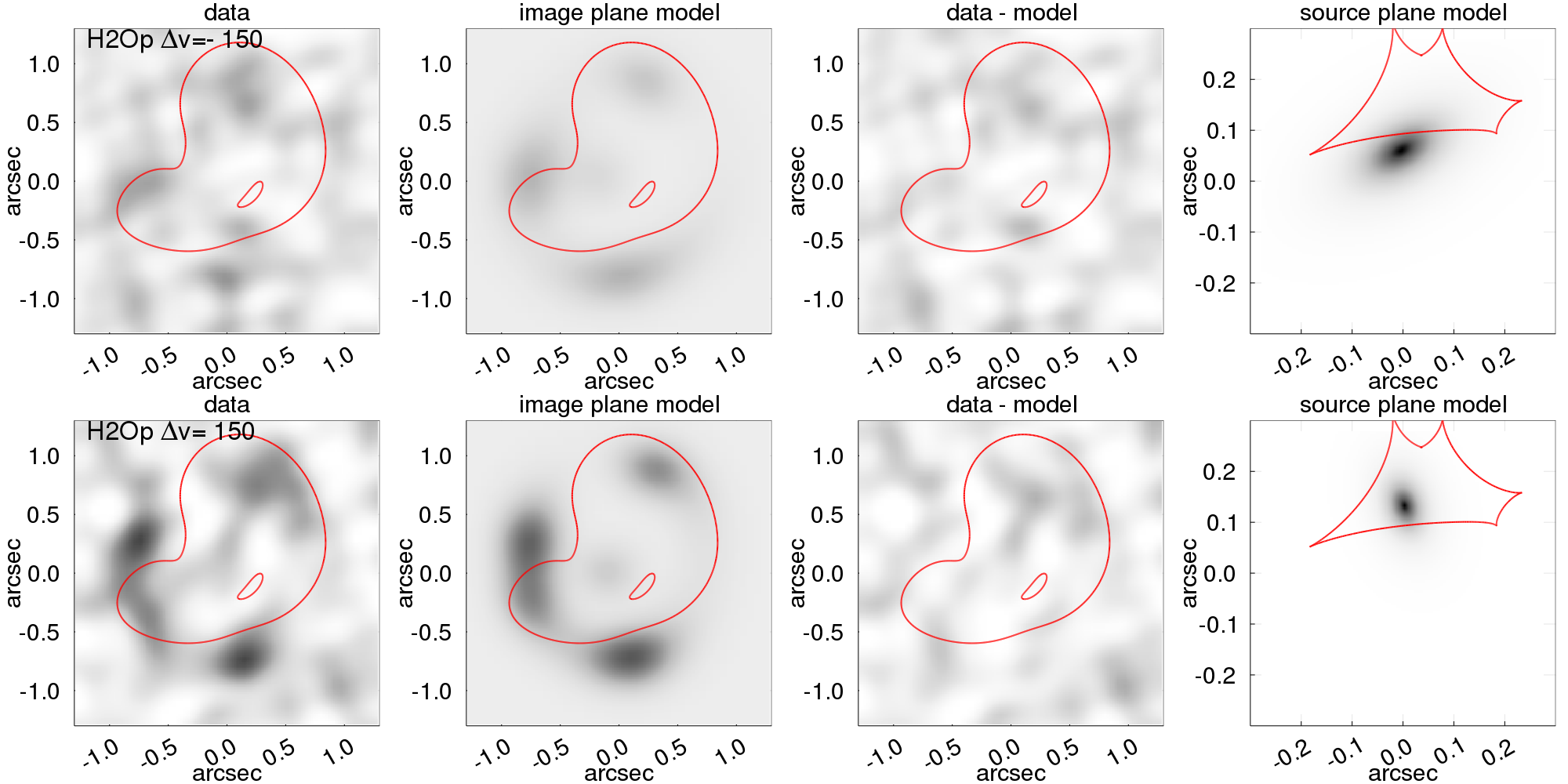}
     \end{center}
     \small{\textbf{Fig.~\ref{fig:allvelchannels}.} Continued, for the \htop\ line whose cube was separated into two bins, collapsed for the entire positive and negative part of the spectrum, respectively.}
 \end{figure*}

\end{CJK*}
\end{document}